\documentclass[manuscript]{acmart}
\acmJournal{TOSEM}

\usepackage{amsmath,amssymb,amsfonts}
\usepackage{algorithmic}
\usepackage{graphicx}
\usepackage{textcomp}
\usepackage{xcolor}
\usepackage[]{hyperref}
\usepackage{xspace}
\usepackage{enumitem}
\usepackage{colortbl}
\usepackage{enumerate}
\usepackage{multirow}
\usepackage{threeparttable}
\usepackage{adjustbox}
\usepackage{lipsum}
\usepackage{hyperref}
\usepackage{makecell}
\usepackage{algorithm}
\usepackage{wrapfig}

\usepackage{tabularx}
\usepackage{array}
\usepackage{adjustbox}
\usepackage{multirow}
\usepackage{balance}
\usepackage{pifont}
\usepackage{booktabs}
\usepackage{tcolorbox}
\usepackage{float}

\usepackage{caption} 

\newcommand{\mytodocomment}[2]{\textcolor{#1}{#2}}

\newcommand{\binchang}[1]{\black{#1}}
\newcommand{\gsz}[1]{\black{#1}}
\newcommand{\yun}[1]{\black{#1}}
\newcommand{\lqy}[1]{\black{#1}}
\newcommand{\zj}[1]{\black{#1}}

\newcommand{\black}[1]{\mytodocomment{black}{{\sf}~#1}}

\newcommand{\graycell}[1]{\colorbox[gray]{0.9}{\makebox[2em][r]{#1}}}
\newcommand{\INDSTATE}[1][1]{\STATE\hspace{#1\algorithmicindent}}
\newcommand{\INDSTATEII}[1][2]{\STATE\hspace{#1\algorithmicindent}}

\AtBeginDocument{%
  }

\begin{document}

\title{Empirical Study of Code Large Language Models for Binary Security Patch Detection}

\author{Qingyuan Li}
\authornote{These authors contribute to the work equally and are co-first authors of the paper.}
\email{522022320072@smail.nju.edu.cn}
\affiliation{
  \institution{National Key Laboratory for Novel Software Technology at Nanjing University}
  \city{Nanjing}
  \country{China}
}

\author{Binchang Li}
\authornotemark[1]
\email{24S151125@stu.hit.edu.cn}
\affiliation{
  \institution{Harbin Institute of Technology}
  \city{Shenzhen}
  \country{China}
}

\author{Cuiyun Gao}
\authornote{Corresponding authors.}
\email{gaocuiyun@hit.edu.cn}
\affiliation{
  \institution{Harbin Institute of Technology}
  \city{Shenzhen}
  \country{China}
}

\author{Shuzheng Gao}
\email{szgao23@cse.cuhk.edu.hk}
\affiliation{
  \institution{Chinese University of Hong Kong}
  \city{Hong Kong}
  \country{China}
}

\author{Zongjie Li}
\authornotemark[2]
\email{zligo@cse.ust.hk}
\affiliation{
  \institution{Hong Kong University of Science and Technology}
  \city{Hong Kong}
  \country{China}
}


\begin{abstract}
Security patch detection (SPD) is crucial for maintaining software security, as unpatched vulnerabilities can lead to severe security risks. 
\lqy{In recent years, numerous learning-based SPD approaches have demonstrated promising results on source code. However, these approaches typically cannot be applied to closed-source applications and proprietary systems that constitute a significant portion of real-world software, as they release patches only with binary files, and the source code is inaccessible.
Given the impressive performance of code large language models (LLMs) in code intelligence and binary analysis tasks such as decompilation and compilation optimization, their potential for detecting binary security patches remains unexplored, exposing a significant research gap between their demonstrated low-level code understanding capabilities and this critical security task.}
\gsz{To address this gap, we construct} a large-scale binary patch dataset containing \textbf{19,448} samples, with two levels of representation: assembly code and pseudo-code, and systematically evaluate \textbf{19} code LLMs of varying scales to investigate their capability in binary SPD tasks.
Our initial exploration demonstrates that directly prompting vanilla code LLMs struggles to accurately identify security patches from binary patches, and even state-of-the-art prompting techniques fail to mitigate the lack of domain knowledge in binary SPD within vanilla models.
\lqy{Drawing on the initial findings, we further investigate the fine-tuning strategy for injecting binary SPD domain knowledge into code LLMs through two levels of representation. Experimental results demonstrate that fine-tuned LLMs achieve outstanding performance, with the best results obtained on the pseudo-code representation.
\binchang{Specifically, models fine-tuned on the pseudo-code dataset outperformed those fine-tuned on the assembly code dataset by an average of \textbf{0.173}, \textbf{0.239}, and \textbf{0.115} in accuracy, F1 score, and false positive rate, respectively.}}
\lqy{To understand the superiority of the pseudo-code representation, we systematically analyze two aspects: embedding features and code naturalness. Embedding distribution of pseudo-code \gsz{exhibits closer alignment with source code compared to assembly code across multiple metrics} with a distance of only \textbf{0.03}, less than one-tenth of that between assembly code and source code. Corroborating the embedding, the code naturalness distribution of pseudo-code aligns more closely with that of source code, where the difference is \textbf{1.307} versus \textbf{4.012} for assembly code. 
These results demonstrate that pseudo-code is more closely aligned with source code, which is the predominant type of code corpus used during LLM pre-training; therefore, pseudo-code is better suited for code LLMs.
Motivated by this finding, we propose a novel augmentation method to enhance the pseudo-code dataset with source code data.
Experimental results demonstrate that code LLMs fine-tuned on the augmented dataset outperform in terms of accuracy (e.g., with a maximum improvement of \textbf{0.147}) and F1 score (e.g., with a maximum improvement of \textbf{0.187}).
}

\end{abstract}



\received{20 February 2007}
\received[revised]{12 March 2009}
\received[accepted]{5 June 2009}

 \maketitle

\section{Introduction}
Timely application of vendor-released security patches is essential, as unpatched vulnerabilities can severely compromise the security of software systems~\cite{frischknecht20171}. 
However, software vendors often release patches without sufficient disclosure~\cite{li2017large}, leaving security patches obscured by numerous non-security patches, thereby challenging the identification of security patches~\cite{mirhosseini2017can}. 
\lqy{This challenge underscores the imperative of automatically identifying security patches.} Consequently, recent research has increasingly focused on security patch detection (SPD)~\cite{lin2024vulnerabilities,wang2023graphspd,tang2023just,wang2021patchrnn}, with most approaches employing deep learning (DL)~\cite{li2023model,samek2021explaining} techniques to analyze syntactic and semantic features in patches for identification. 

Current learning-based approaches for SPD in source code can be categorized into \lqy{Sequence-based, Graph-based, and LLM-embedding-based methods}. 
Sequence-based methods utilize sequence neural networks to capture semantic information from all sequential inputs. For example, PatchRNN~\cite{wang2021patchrnn} employs a recurrent neural network (RNN)~\cite{zaremba2014recurrent} by using code changes and commit messages as input to identify security patches. 
Graph-based methods utilize graph neural networks (GNNs)~\cite{ggnn} to capture graph information from code changes. For example, GraphSPD~\cite{wang2023graphspd} proposes a PatchCPG as input and uses a graph convolutional network (GCN)~\cite{kipf2016gcn} to convert PatchCPG into graph representations to detect security patches.
LLM-embedding-based methods utilize large language models (LLMs) to encode code changes and other messages into semantic embeddings. For example, LLMDA~\cite{tang2023just} leverages CodeT5~\cite{wang2021codet5} and CodeLlama~\cite{codellama} as embedding layers and trains an attention-based network to align these different embeddings to determine whether a code commit fixes a vulnerability.
\zj{Although many SPD methods have been proposed for open-source software (OSS), a significant portion of real-world software ecosystems consists of closed-source applications and proprietary systems~\cite{economides2006two,lee2005open}, where patches are distributed only as binary files without accessible source code. As binary code represents a fundamentally different abstraction level from source code, existing approaches targeted at source code cannot be directly applied to the binary scenario, posing significant challenges to SPD tasks on closed-source software and thereby threatening software security in these critical environments.}

\zj{Studies focusing on binary SPD remain limited and face substantial technical challenges, }\lqy{such as the paucity of publicly available binary datasets and the semantic gap between binary code and source code.}
\zj{\lqy{The limited body of existing} studies can be broadly categorized into Pattern-based, Rule-based, and Learning-based methods.} 
SPAIN~\cite{xu2017spain} is a pattern-based method for binary-level patch analysis. It identifies changed traces on assembly-level control flow graphs (CFGs) to summarize patch patterns and corresponding vulnerability patterns.
1dVul~\cite{peng20191dvul} is a rule-based method that detects binary security patches using heuristic rules. It fed fuzzing test inputs to the unpatched program to confirm if the patch is security-related.
BinGo~\cite{he2024bingo} is a learning-based method that constructs code property graphs derived from CFGs and applies GNNs to detect security patches.
Despite the efforts, these methods trade off between time-consuming dynamic analyses and model complexity, yet still yield limited accuracy.
\zj{\lqy{Concurrently}, large language models (LLMs) have revolutionized various software engineering tasks, particularly in understanding and processing code representations.} 
\zj{Recent advances have demonstrated LLMs' remarkable capabilities in \lqy{several low-level code analysis tasks}, including decompilation~\cite{tan2024llm4decompile}, compilation optimization~\cite{cummins2025llm}, and assembly code understanding~\cite{jiang2025nova}. These models, pre-trained on extensive code corpora and further pre-fine-tuned on low-level representations, have shown promising \lqy{capabilities in bridging} the semantic gap between high-level source code and low-level binary representations. Despite these encouraging developments, the potential of code LLMs for binary security patch detection remains unexplored, presenting a significant research opportunity to leverage their demonstrated low-level code understanding capabilities for this critical security task.}

\lqy{In this paper, we conduct the first empirical study to systematically evaluate the capabilities of code LLMs on binary SPD tasks.}
\lqy{Given the absence of existing datasets for binary SPD, we construct a large-scale binary SPD dataset. Inspired by the prior study~\cite{he2024bingo}, we select five representative open-source projects, including Linux, FFmpeg, Git, PHP, and Libav, from two widely used source-code-level SPD datasets, ReposVul~\cite{wang2024reposvul} and PatchDB~\cite{wang2021patchdb}. We subsequently automatically compile the corresponding pre- and post-patch source files from these projects to generate binary artifacts for our study. After the compilation process, we lift the collected binaries into two different abstraction levels: 1) Assembly-code representation via disassembly, and 2) Pseudo-code representation via decompilation.
After post-processing, the resulting corpus contains \textbf{19,448} labeled samples, comprising \textbf{8,311} security patches and \textbf{11,137} non-security patches.
We further evaluate 19 representative off-the-shelf code LLMs of varying sizes (ranging from 0.5B to 9B) as well as two powerful foundation models via three widely used prompting strategies, including zero-shot~\cite{zhao2023pre}, chain-of-thought (CoT)~\cite{kojima2022large}, and self-correction~\cite{madaan2023self}, to assess whether they can be directly applied to the binary SPD task. 
Experimental results demonstrate that all evaluated models exhibit limited performance across both representations of binary patches (i.e., assembly code and pseudo-code). Although CoT and self-correction provide improvements in certain models compared to zero-shot, they remain unsatisfactory in identifying security patches. This suggests that these prompting strategies fail to mitigate the lack of domain knowledge in binary SPD within off-the-shelf models.
}
\yun{
Considering that fine-tuning has been widely recognized as an effective strategy for adapting vanilla LLMs to specialized downstream tasks.
\lqy{To investigate whether vanilla code LLMs can be adapted to binary SPD tasks, we further explore the effectiveness of fine-tuning code LLMs via Low-Rank Adaptation (LoRA)~\cite{hu2022lora}, one of the most widely used techniques for efficient model fine-tuning.}
Experimental results demonstrate that code LLMs perform well on the pseudo-code dataset, achieving much higher performance than that on the assembly-code dataset.
} 
Moreover, fine-tuned code LLMs exhibit robustness across different optimization levels on the pseudo-code dataset.
\lqy{
To unveil the superiority of the pseudo-code representation in binary SPD tasks, we systematically analyze the correlation between assembly code, pseudo-code, and their corresponding source code, utilizing two widely adopted approaches, embeddings and code naturalness, to represent code features~\cite{utpala2023language,wehrmann2019language,hindle2016naturalness,yang2024dependency}.
Experimental results reveal that pseudo-code aligns more closely with source code. Since code LLMs perform best on source-code-related tasks~\cite{lozhkov2024starcoder,deepseekai2025deepseekr1incentivizingreasoningcapability,deepseekcoderv2}, pseudo-code is a more suitable data representation for binary SPD tasks.
}
\lqy{
Motivated by this finding, we propose a novel augmentation method to enhance the binary SPD dataset.
This method incorporates source code data to augment our binary dataset of the pseudo-code representation, leveraging the similarity between pseudo-code and source code. We fine-tune 19 code LLMs on the augmented dataset. 
Experimental results demonstrate that this augmentation method further improves model performance, with more pronounced effects observed on smaller-scale models.
}

In summary, the \textbf{main contributions} of this paper are concluded as follows:

(1) To the best of our knowledge, we represent the first systematical study on applying code LLMs to binary SPD tasks.

(2) We develop the first open-source, large-scale, multi-project binary security patch detection dataset spanning multiple compiler optimization levels, and comprehensively evaluate 19 code LLMs \lqy{through prompting and fine-tuning}.

\lqy{(3) We identify pseudo-code representation as the best-suited data format for applying code LLMs to the binary SPD task, and further investigate the underlying reasons for its effectiveness.}

\lqy{(4) We propose a novel augmentation method to enhance the binary SPD dataset and validate its effectiveness by fine-tuning 19 code LLMs on the augmented dataset.}

(5) We publicly disclose our binary SPD dataset, fine-tuned code LLMs, and the source code at \url{https://github.com/anonauthors123/LLM-for-Binary-SPD}.

\gsz{\textbf{Organization.} The rest of this paper is organized as follows:} 
\binchang{Section~\ref{sec:background} introduces the background of binary security patch security detection and gives a motivating example of our study. Section~\ref{sec:study-design} formulates research questions and describes the design of our study. Section~\ref{sec:results-analysis} analyzes the experiment results and answers the research questions. Section~\ref{sec:error-analysis} delves into LLMs' error examples to provide more detailed insights. Section~\ref{sec:threats-to-validity} discusses the potential threats to validity in our study. Section~\ref{sec:related-work} introduces previous works in several related topics. And Section~\ref{sec:clucusion} concludes the paper.}

\section{Background}
\label{sec:background}
In this section, we review representative approaches to binary security patch detection and discuss their limitations. We then articulate our research motivation through an illustrative example.

\subsection{Binary Security Patch Detection}
In some scenarios, developers cannot access the source code patches in software projects (e.g., updates for closed‐source software). 
In such cases, only binary artifacts are available for identifying security patches. 
Given the limited research on binary security patch detection (SPD), we categorize prior work into three representative technical directions.

\textbf{\textit{Rule-based}}. 1dVul~\cite{peng20191dvul} is a binary security patch discovery approach based on heuristic rules and symbolic execution. 
It first identifies target branches in binary patches using heuristic rules, then employs a distance-based fuzzing mechanism to prioritize test inputs closer to the target. Finally, it applies a symbolic execution to generate inputs that reach the target branch and verify the security relevance of patches.

\textbf{\textit{Pattern-based}}. SPAIN~\cite{xu2017spain} proposes a pattern-based approach for binary-level patch analysis. It locates patched functions, identifies changed traces on assembly-level control flow graphs (CFGs), and performs semantic analysis on the changed traces to summarize patch patterns and corresponding vulnerability patterns for security patch identification.

\textbf{\textit{Learning-based}}. PMatch~\cite{lang2021pmatch} employs a CBOW neural
network model to generate the semantic feature vectors of assembly instructions, thereby identifying security patches. BinGo~\cite{he2024bingo} employs a graph neural network (GNN) over code property graphs derived from CFGs, to capture both structural and semantic characteristics of binary code, enabling accurate identification of security patches.

\subsection{Motivating Example}
Existing approaches suffer from several limitations.
Rule‐based approaches rely on complex symbolic analysis~\cite{peng20191dvul}.
Pattern‐based approaches suffer from low accuracy~\cite{xu2017spain}. 
Additionally, learning-based approaches require training from scratch~\cite{he2024bingo,lang2021pmatch}, and sequence-based networks (e.g., the CBOW network) are limited by the length of assembly instructions~\cite{lang2021pmatch}.
However, emerging studies have revealed that in certain software engineering (SE) tasks, techniques leveraging large language models (LLMs) surpass rule-based and pattern-based approaches~\cite{li2023nuances,thapa2022transformer,mastropaolo2021empirical,xia2022less,zhang2023gamma}.
Meanwhile, the extended long-context capabilities of LLMs largely alleviate the constraints associated with input length limitations~\cite{an2024make,jin2024llm}.
The remarkable achievements of LLMs in SE tasks motivate us to investigate their capabilities to address binary SPD. However, directly disassembling binary patches into assembly-code representations may not be suitable for code LLMs.

\begin{wrapfigure}{r}{0.5\textwidth} 
\centering
\vspace{-1em}
\includegraphics[width=0.9\linewidth]{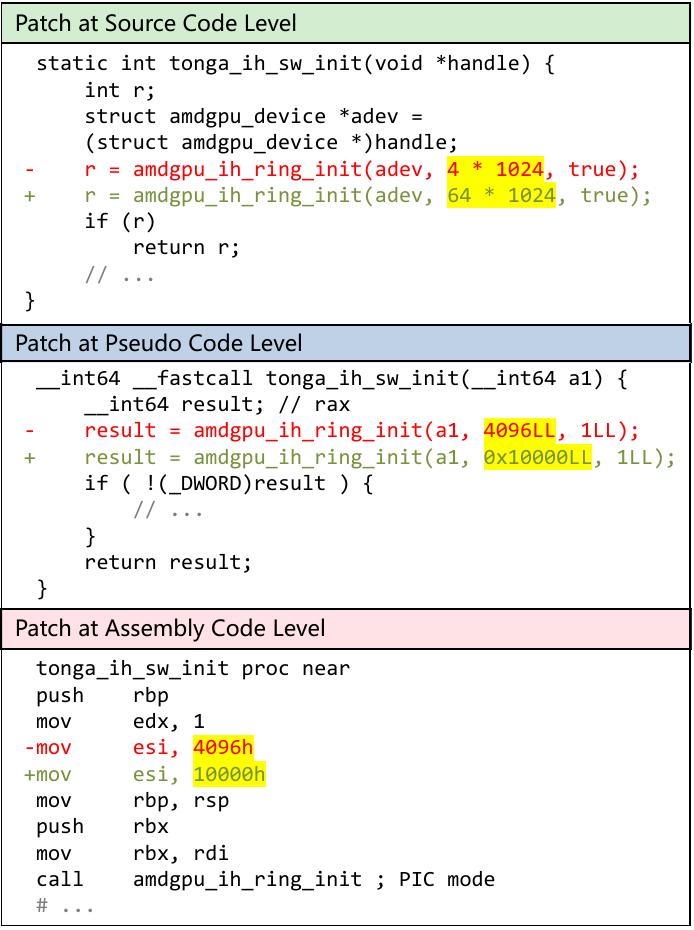}
\caption{A motivating example from Linux Kernel, and the highlighted parameter is the IH ring buffer size}
\label{fig:motivation}
\vspace{-3em}
\end{wrapfigure}

Figure~\ref{fig:motivation} illustrates a security patch corresponding to the Linux kernel. This patch increases the Interrupt Handler (IH) ring buffer size to avoid overflow.
Given the three patch forms (i.e., source, pseudo, and assembly code patches), pseudo code effectively restores the original information, including conditional statements and variable types. In contrast, assembly code transforms into register operations, losing almost all structural and semantic information inherent to high-level code.

The code changes further reveal this situation.
In the pseudo code, the code change accurately restores the function call and parameter passing of the \textit{amdgpu\_ih\_ring\_init}, achieving syntactic and semantic alignment with the source code.
Conversely, in the assembly code, it manifests as a three-address code (3AC) where the hexadecimal value is loaded into the \textit{esi} register.
The function call and parameter passing of \textit{amdgpu\_ih\_ring\_init} are scattered in the subsequent context, leaving only parameter modifications, thus dispersing the information in the code change.
Building upon the motivating example, we systematically investigate two detection settings for leveraging code LLMs in binary SPD: pseudo code versus assembly code.


\section{Study Design}
\label{sec:study-design}
In this section, we detail the systematic study, including the research questions (RQs), the construction of the binary security patch detection (SPD) dataset, the evaluated code LLMs, the experimental environments, the evaluation metrics, and the experimental procedure.

\begin{figure*}[h]
\centering
\includegraphics[width=0.9\linewidth]{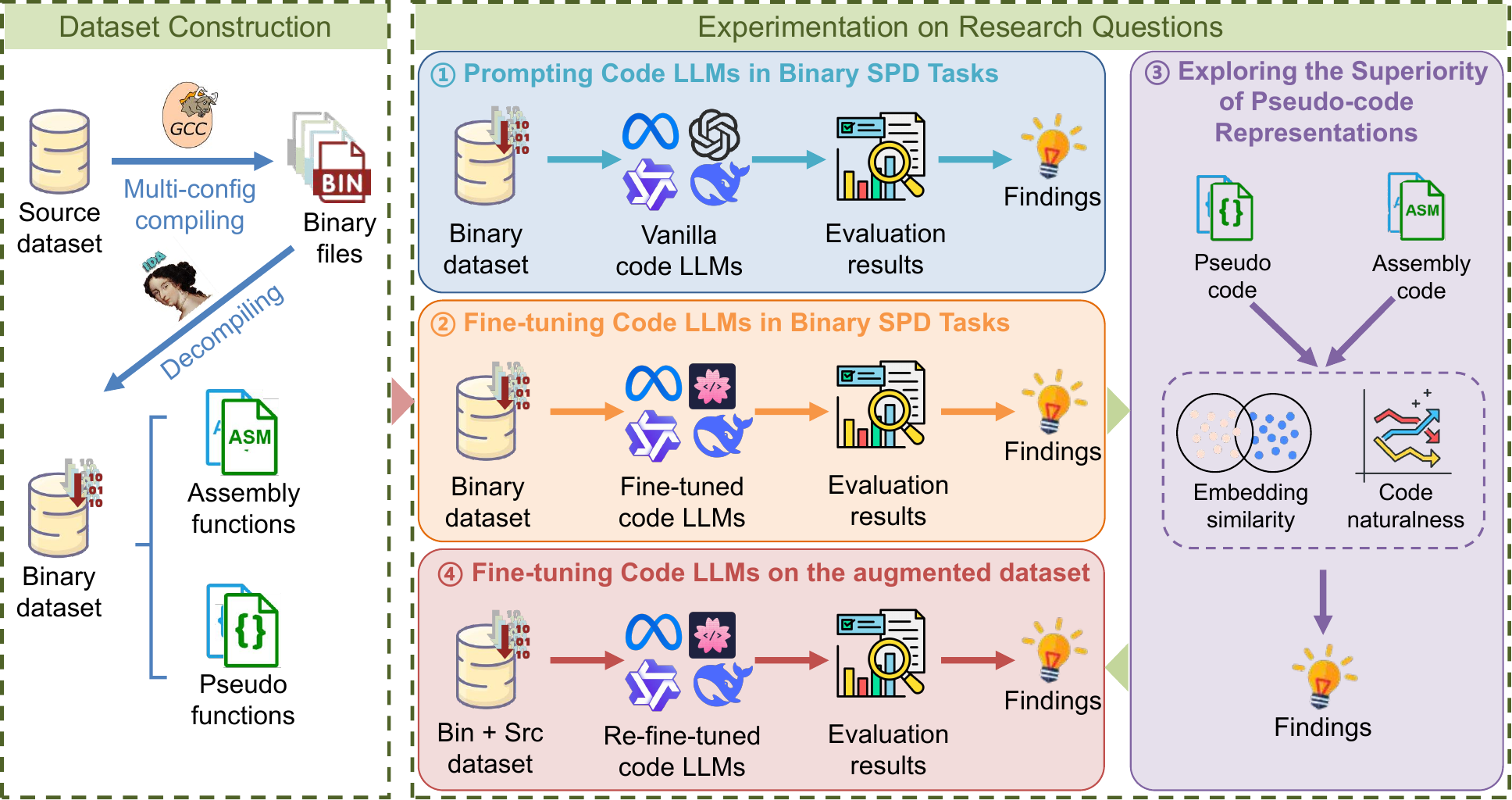}
\caption{Overview of the empirical study}
\label{fig:overview}
\vspace{-1em}
\end{figure*}

Figure~\ref{fig:overview} is the overview of our study. Given the absence of datasets for binary SPD, we first construct a large-scale binary SPD dataset as illustrated in the \textbf{Dataset construction} part. 
\lqy{We utilize three widely used prompting strategies that enhance model reasoning to evaluate code LLMs on the newly constructed binary SPD dataset to assess the effectiveness of directly prompting for binary SPD tasks, as detailed in the \textbf{RQ$\textcircled{1}$} part. We further employ Low-Rank Adaptation (LoRA)~\cite{hu2022lora}, one of the most widely used fine-tuning techniques, to fine-tune code LLMs on this dataset to assess the effectiveness of fine-tuning for binary SPD tasks, as detailed in the \textbf{RQ$\textcircled{2}$} part. Based on the superior performance of fine-tuned code LLMs on the pseudo-code representation, we conduct a comparative study to investigate the superiority of the pseudo-code representation. Specifically, we compare pseudo-code, assembly code, and source code along two aspects, embedding characteristics and code naturalness, to determine whether pseudo-code more closely aligns with source code and is therefore better suited for code LLMs, as detailed in the \textbf{RQ$\textcircled{3}$} part. Motivated by the observed similarity between pseudo-code and source code, we perform source code data augmentation on the binary dataset and further fine-tune the code LLMs to investigate whether augmenting with source code data improves model effectiveness on binary SPD tasks, as detailed in the \textbf{RQ$\textcircled{4}$} part.}

\subsection{Research Questions}
We investigate the effectiveness of code LLMs in detecting binary security patches by answering the following research questions (RQs).

\lqy{
\textbf{RQ1}: Can code LLMs be directly applied to binary security patch detection via prompting?}

\lqy{
\textbf{RQ2}: Can code LLMs be effectively adapted to binary security patch detection via fine-tuning?}

\lqy{
\textbf{RQ3}: Why do fine-tuned code LLMs demonstrate superior performance on pseudo-code representations?}

\lqy{
\textbf{RQ4}: Does augmenting the binary dataset with source code data enhance the effectiveness of model fine-tuning?
}

\subsection{Dataset Construction}
\begin{figure}[htp]
\centering
\includegraphics[width=0.85\linewidth]{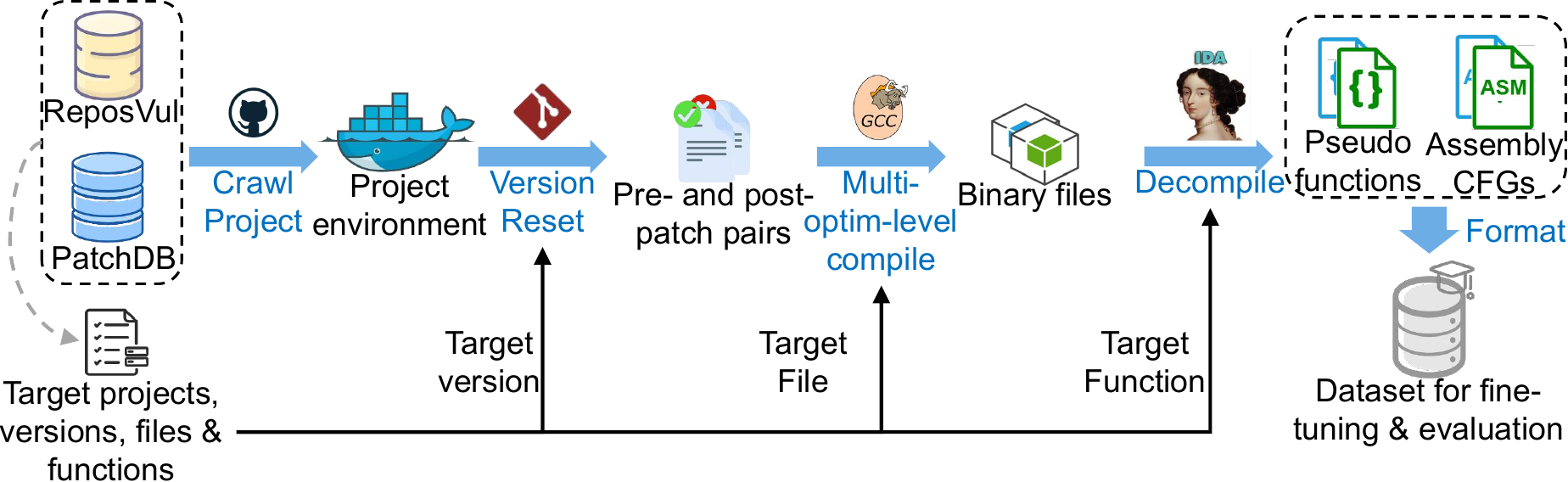}
\vspace{-1em}
\caption{Construction workflow of the binary SPD dataset}
\label{fig:construct_workflow}
\vspace{-1em}
\end{figure}

In this paper, we construct a large-scale, multi-project, and multi-optimization-level binary security patch detection (SPD) dataset. 
\lqy{Figure~\ref{fig:construct_workflow} presents the data collection process, which comprises four steps:
1) Identifying the target compilation projects, 
2) compiling the  source files into binaries, 
3) transforming binary pre- and post-patch pairs into assembly-code and pseudo-code representations, and 
4) filtering and splitting the dataset. 
We describe them in detail as follows.}

\textbf{\textit{Projects Selection}}. 
\lqy{Inspired by~\citet{he2024bingo}, we construct the binary SPD dataset based on two popular pre-collected source-code patch datasets, ReposVul~\cite{wang2024reposvul} and PatchDB~\cite{wang2021patchdb}. 
PatchDB represents one of the most frequently used datasets in source-code SPD, and ReposVul ensures patch quality through a dual-validation mechanism.
We enumerate all projects containing patches in both datasets and rank them by the number of patches and project popularity.
A higher patch count ensures that compiling a project yields more binaries, while popularity serves as an indicator of project quality.
We then select projects with more than 1,000 GitHub stars and at least 100 patches, followed by manually verifying that they can automatically compile the pre- and post-patch files within the project.}

\textbf{\textit{Source Files Compilation}}. 
\lqy{
To reduce compilation time, we compile only the pre- and post-patch source files located in subdirectories at multiple optimization levels. We manually prepare the building environments for different projects and employ the GCC compiler to compile all source code files, as it is the most widely used C/C++ compiler. }
We design an automated compilation methodology, as shown in Algorithm~\ref{alg:compiling}. 
The process begins with \textit{GetCommitIds}, which retrieves all patch versions for a specific project from both datasets. Subsequently, \textit{GetFileNames} extracts pre- and post-patch filenames from commit messages. The \textit{GitReset} executes the \textit{git reset -{}-hard} command to rollback projects to target versions: first to the patched version, then to the preceding defective version using \textit{HEAD\^}. The \textit{MakeDefConfig} handles compilation configuration. The \textit{MultiOptimizationLevelCompile} function is the core component of the compilation algorithm. 
The design of the \textit{MultiOptimizationLevelCompile} function is as follows: 
First, given a target source file (either a .c file or a .cpp file), the function checks whether a Makefile exists in the file's subdirectory. 
If so, the function modifies the compilation optimization level, changing it sequentially from O0 to Os, and compiles the source file for each optimization level. 
After compilation, the resulting binary file (a .o file) is renamed and stored.

\begin{algorithm}[htb]
\caption{Automated compiling defect-patch files}
\renewcommand{\algorithmicrequire}{\textbf{Input:}}
\renewcommand{\algorithmicensure}{\textbf{Output:}}
\begin{algorithmic}[1]
\label{alg:compiling}
\REQUIRE Source project $P_i$, Primary datasets $D_{src}$
\ENSURE Binary files $F_{bin}$

    \STATE // Automated compile files with multi-optimization levels
    \STATE \textbf{def} MultiOptimizationLevelCompile($file$):
        \INDSTATE $file\_list \gets []$
        \INDSTATE $file\_dir \gets \text{GetParentDirectory}(file)$
        \INDSTATE \textbf{if} $\text{Exists}(file\_dir)$ \ \textbf{then}
            \INDSTATEII $makefile \gets \text{GetMakeFile}(file\_dir)$
        \INDSTATE \textbf{end if}
        \INDSTATE $optim\_levels \gets [O0, O1, O2, O3, Os]$
        \INDSTATE \textbf{for} each $level \in optim\_levels$
            \INDSTATEII $\text{SetCompilerFlags}(level)$
            \INDSTATEII $file_{bin} \gets \text{CompileFile}(file)$
            \INDSTATEII $file_{renamed} \gets \text{FormatRename}(file_{bin})$
            \INDSTATEII $file\_list\text{.append}(file_{renamed})$
        \INDSTATE \textbf{end for}
        \INDSTATE \textbf{return} $file\_list$
    \STATE \textbf{end def}

    \STATE $commit\_ids \gets \text{GetCommitIds}(P_i, D_{src})$
    \STATE $F_{bin} \gets []$
    \FOR{each $id \in commit\_ids$}
        \STATE $\text{GitReset}(id)$
        \STATE $\text{MakeDefConfig}()$
        \STATE $file\_names \gets \text{GetFileNames}(id)$
        \FOR{each $file \in file\_names$}
            \STATE // Compile the patched file
            \STATE $files_{bin} \gets \text{MultiOptimizationLevelCompile}(file)$
            \STATE $F_{bin}\text{.extend}(files_{bin})$
        \ENDFOR
        \STATE // Rollback the project to the previous version
        \STATE \text{GitReset}(\textit{HEAD\^})
        \STATE $\text{MakeDefConfig}()$
        \FOR{each $file \in file\_names$}
            \STATE // Compile the defective file
            \STATE $files_{bin} \gets \text{MultiOptimizationLevelCompile}(file)$
            \STATE $F_{bin}\text{.extend}(files_{bin})$
        \ENDFOR
    \ENDFOR
    \STATE $\text{CopyFiles}(F_{bin}, tgt\_dir)$
    \RETURN $F_{bin}$
\end{algorithmic}
\end{algorithm}


\begin{figure}[htbp]
  \centering
  \begin{minipage}[b]{0.48\textwidth}
    \centering
    \includegraphics[width=0.93\linewidth]{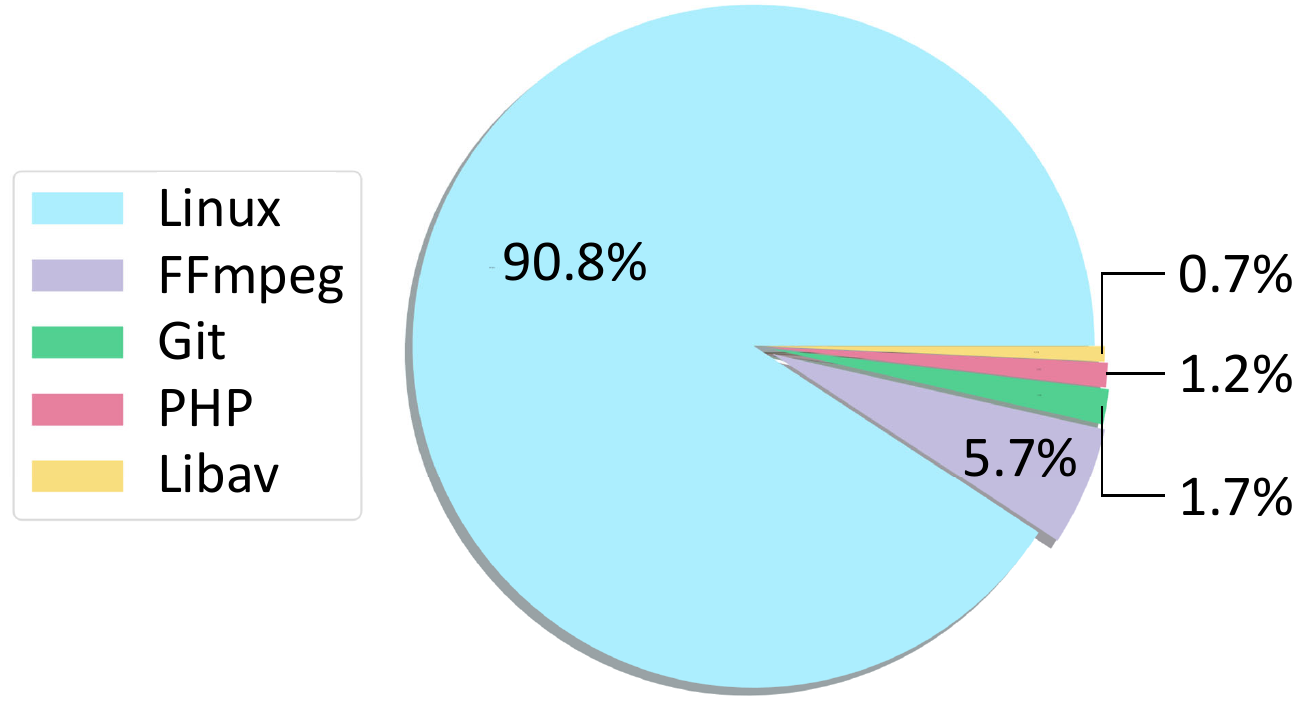}
    \caption{Pie chart of source projects' distribution}
    \label{fig:proj_distr}
  \end{minipage}
  \hfill
  \begin{minipage}[b]{0.48\textwidth}
    \centering
    \includegraphics[width=0.7\linewidth]{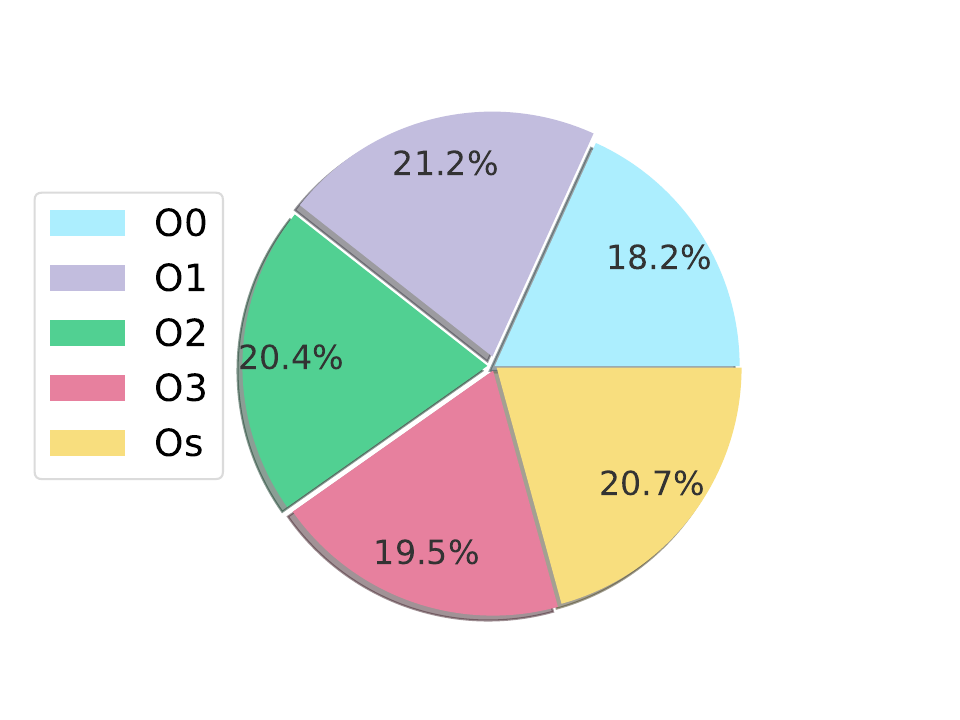}
    \caption{Pie chart of optimization levels' distribution}
    \label{fig:optim_distr}
  \end{minipage}
\end{figure}

\textbf{\textit{Binaries Decompilation}}. 
We employ IDA Pro\footnote{\url{https://hex-rays.com/ida-pro}.} to disassemble binaries into \textbf{assembly code} and decompile them into \textbf{pseudo code}. Function names for pre- and post-patch versions are extracted from corresponding commit messages. We then construct function-level datasets containing assembly control flow graphs (CFGs) and pseudo functions.

\lqy{
\textbf{\textit{Data Post-processing}}.
We first perform commit-level deduplication to ensure that data from the same commit appears only once. 
Subsequently, a double-threshold filter is applied to remove functions that are excessively short or long, improving the feasibility of subsequent evaluation. 
The data of \textbf{assembly CFGs} and \textbf{pseudo functions} is then converted into the Alpaca format and split into training, validation, and test sets with an 8:1:1 ratio for fine-tuning experiments. 
As a final step, exact whole-word matching is applied across the splits to ensure that no identical items appear in more than one set.
}

Finally, the dataset contains 19,448 binary pre- and post-patch pairs, comprising 8,311 security patches and 11,137 non-security patches. 
Figure~\ref{fig:proj_distr} visualizes the relative contribution of each source project to the overall dataset composition, and Figure~\ref{fig:optim_distr} illustrates different optimization level proportions.



\subsection{Experimental Setup}
\textbf{\textit{Selected LLMs}}. To comprehensively evaluate the ability to detect binary security patches across different model scales, we conduct experiments with nineteen open-source code LLMs, ranging from 0.5B to 9B parameters, as well as two proprietary state-of-the-art foundation models.
We select seven well-proven code model series in software engineering, including CodeLlama~\cite{codellama}, Qwen2.5-Coder~\cite{qwen2.5coder}, DeepSeek-Coder~\cite{deepseekllm, deepseekcoderv2}, Yi-Coder~\cite{yi-coder}, OpenCoder~\cite{opencoder}, StarCoder2~\cite{lozhkov2024starcoder}, aiXcoder~\cite{jiang2024aixcoder}, and two emerging binary code model series, including LLMCompiler~\cite{cummins2025llm} and LLM4Decompile~\cite{tan2024llm4decompile}.
Additionally, we select GPT-3.5-Turbo~\cite{achiam2023gpt} and DeepSeek-R1~\cite{deepseekai2025deepseekr1incentivizingreasoningcapability} as foundation models, both of which have demonstrated stable performance across a wide range of code intelligence tasks.

\textbf{\textit{Experimental environments}}. We run experiments on a server with Ubuntu OS, which is equipped with a 96-core CPU (Intel Xeon Platinum 8260) and four GPUs (NVIDIA A100 Tensor Core GPU 40GB) with CUDA version 12.2.

\textbf{\textit{Evaluation metrics}}. We choose the following metrics to evaluate the performance of code LLMs.
\begin{itemize}
    \item \textbf{Accuracy:} $Accuracy=\frac{TP+TN}{TP+TN+FN+FP}$. This metric quantifies the percentage of the samples that are correctly classified among all samples. We denote patches that fix vulnerabilities (security patches) as positive samples, and those do not as negative samples. $TP$ and $TN$ denote the counts of true positive and true negative samples, respectively. $TP+TN+FN+FP$ represents the total number of samples.


    
    \item \textbf{F1 score:} $F1=\frac{2\times Precision\times Recall}{Precision+Recall}$. This metric is the harmonic mean of precision and recall.
     
    \item \textbf{False Positive Rate (FP Rate):} $FPR=\frac{FP}{FP+TN}$. This metric measures the proportion of negative samples that are erroneously classified as positive. 
     
\end{itemize}

\subsection{Experimental Procedure}

\lqy{To answer \textbf{RQ1}, we employ three widely used prompting strategies to enhance model reasoning, including zero-shot~\cite{zhao2023pre}, chain-of-thought (CoT)~\cite{kojima2022large}, and self-correction~\cite{madaan2023self}, to evaluate the capability of 19 code LLMs and two proprietary foundation models when directly applied to binary SPD. To maximize models' instruction-following to obtain a definitive judgment from code LLMs, we require the models to perform chain-of-thought (CoT) reasoning and self-correction implicitly, and to respond only with ``yes'' or ``no''.
Inference hyperparameters of all models are fixed as follows: \textit{temperature=0} and \textit{top\_p=0} to minimize output randomness and eliminate sampling-induced perturbations.}

\lqy{To answer \textbf{RQ2}, we fine-tune 19 code LLMs on datasets at two abstraction levels, assembly-code and pseudo-code, to investigate 1) whether fine-tuning yields significant improvements in binary SPD, and 2) which data representation (pseudo-code vs. assembly-code) is more effective for model fine-tuning. 
We adopt Low-Rank Adaptation (LoRA)~\cite{hu2022lora} as our fine-tuning method, leveraging its parameter efficiency to substantially reduce computational and memory costs while maintaining competitive performance~\cite{hu2022lora}. Given its wide adoption in both natural language and code-related tasks, LoRA provides a strong and practical choice for evaluating the effectiveness of fine-tuning. 
We conduct all experiments using the LLaMAFactory framework\footnote{\url{https://github.com/hiyouga/LLaMA-Factory}}, with the following training hyperparameters: \textit{num\_train\_epochs=3}, \textit{learning\_rate=1.0e-4}, \textit{batch\_size=2}, \textit{cutoff\_len=4096}, \textit{lora\_rank=8}, and \textit{lora\_alpha=16}.}

\lqy{To answer \textbf{RQ3}, we first posit two assumptions: 1) current code LLMs are trained predominantly on source code, and 2) LLMs perform better on languages that constitute a larger share of their pretraining corpora. 
Evidence that modern code-specialized models rely primarily on source-code corpora, which center pretraining on large public source-code archives~\cite{lozhkov2024starcoder,deepseekai2025deepseekr1incentivizingreasoningcapability,deepseekcoderv2}. Furthermore, empirical research demonstrates that increasing the proportion of code in pretraining mixtures improves model performance on structured and code-related tasks~\cite{li2024quantifying,petty2025does}.
From these assumptions, we infer that code representations that are closer to source-code representations yield better performance for code LLMs. Accordingly, we compare the embedding characteristics and the code naturalness of assembly code and pseudo-code with those of source code to assess the similarity among the three representations and thereby determine whether pseudo-code exhibits greater semantic and structural proximity to source code. Embedding characteristics have been used to quantify the semantic properties of programming languages~\cite{utpala2023language,wehrmann2019language,ahmed2024studying}, while code naturalness, operationalized via language-model-based predictability measures and their dependency-aware extensions, captures repetitiveness and predictability in code and thus partly reflects structural properties~\cite{hindle2016naturalness,yang2024dependency}.}

To compare the similarity of assembly code and pseudo-code to those of source code in embedding characteristics, we employ a code embedding model, jina-embeddings-v2-base-code\footnote{\url{https://huggingface.co/jinaai/jina-embeddings-v2-base-code}}, to obtain the embedding vectors of all code snippets. This embedding model supports multiple programming languages, including Assembly, C, and C++. We then leverage t-SNE~\cite{van2008visualizing} to project the embedding vectors into a normalized two-dimensional space and visualize the projection with a scatter plot. Finally, we compute the distance between the centroids of different code representations. 
To compare the similarity of assembly code and pseudo-code to those of source code in code naturalness, we measure code naturalness using cross-entropy, a widely adopted metric for this purpose~\cite{ray2016naturalness,hindle2016naturalness}. Equation~\ref{con:cross_entropy} represents the cross-entropy of a code snippet, where $S = y_{1} \dots y_{N}$ denotes a code snippet of length $N$ and $y_{i}$ represents an individual token in the code snippet. A language model $M$ can be represented as a probability distribution $p(\cdot)$. We utilize the CodeT5-large~\cite{wang2021codet5} to measure the code naturalness, as it has been used frequently in previous studies~\cite{yang2023large,xia2023automated}. Owing to the large scale of the datasets, we calculate the code naturalness by sampling one-tenth of the assembly, pseudo, and source code, respectively.
\begin{equation}
    \mathit{CE} = -\frac{1}{N} \sum_{i=1}^{N} \log_{}{p(y_{i})}
    \label{con:cross_entropy}
\end{equation}


\lqy{To answer \textbf{RQ4}, we collect an additional source-code SPD dataset from ReposVul~\cite{wang2024reposvul} and PatchDB~\cite{wang2021patchdb} to augment the pseudo-code dataset. The source-code dataset comprises 23,222 patches and applies the same commit-level deduplication and double-threshold filtering procedures used when constructing the binary dataset, thereby ensuring dataset quality.
We employ identical fine-tuning techniques and hyperparameter configurations as those used in RQ2 to fine-tune all 19 language models. 
To investigate effective data augmentation strategies, we design two different approaches: 1) direct dataset merging (while maintaining pseudo-code data exclusively in validation and test sets), and 2) sequential fine-tuning that first utilizes the source-code dataset, followed by the pseudo-code dataset. 
To find a more suitable strategy, we conduct a preliminary experiment on 0.5B and 1.5B parameters LLMs with both strategies.
The experiment results show that all tested LLMs perform better with the first strategy. Following the results, we subsequently apply the direct dataset merging strategy for fine-tuning the remaining LLMs.}

\section{Results Analysis}
\label{sec:results-analysis}

\subsection{Answer to RQ1: Prompting Yields Only Limited Effectiveness for Code LLMs on Binary SPD Tasks}

\begin{table*}[htbp]
\centering
\caption{Evaluation results of prompting code LLMs on binary SPD}
\label{tab:tab1}
\scriptsize 
\renewcommand{\arraystretch}{0.9} 
\setlength{\tabcolsep}{6pt} 

\newcommand{\scaleA}[1]{\pgfmathparse{1*#1}\pgfmathprintnumber[fixed,precision=3]{\pgfmathresult}} 
\newcommand{\scaleB}[1]{\pgfmathparse{1*#1}\pgfmathprintnumber[fixed,precision=3]{\pgfmathresult}} 
\newcommand{\scaleC}[1]{\pgfmathparse{1*#1}\pgfmathprintnumber[fixed,precision=3]{\pgfmathresult}} 

\ExplSyntaxOn
\NewDocumentCommand{\OneRowModel}{mm}{
  #1 & Zero-shot
    & \clist_item:nn {#2} {1} & \clist_item:nn {#2} {2}
    & \clist_item:nn {#2} {3} & \clist_item:nn {#2} {4}
    & \clist_item:nn {#2} {5} & \clist_item:nn {#2} {6}
    & \clist_item:nn {#2} {7} & \clist_item:nn {#2} {8} \\
}
\ExplSyntaxOff

\ExplSyntaxOn
\NewDocumentCommand{\ThreeRowModelFull}{mmmm}
 {
  \multirow{3}{*}{#1} & Zero-shot
    & \clist_item:nn {#2} {1} & \clist_item:nn {#2} {2}
    & \clist_item:nn {#2} {3} & \clist_item:nn {#2} {4}
    & \clist_item:nn {#2} {5} & \clist_item:nn {#2} {6}
    & \clist_item:nn {#2} {7} & \clist_item:nn {#2} {8} \\
  & CoT
    & \clist_item:nn {#3} {1} & \clist_item:nn {#3} {2}
    & \clist_item:nn {#3} {3} & \clist_item:nn {#3} {4}
    & \clist_item:nn {#3} {5} & \clist_item:nn {#3} {6}
    & \clist_item:nn {#3} {7} & \clist_item:nn {#3} {8} \\
  & Self-corr
    & \clist_item:nn {#4} {1} & \clist_item:nn {#4} {2}
    & \clist_item:nn {#4} {3} & \clist_item:nn {#4} {4}
    & \clist_item:nn {#4} {5} & \clist_item:nn {#4} {6}
    & \clist_item:nn {#4} {7} & \clist_item:nn {#4} {8} \\
  \addlinespace[0.6ex]
 }
\ExplSyntaxOff

\resizebox{\textwidth}{!}{%
\begin{threeparttable}
\begin{tabular}{cc|cccc|cccc}
\toprule
\multirow{2}{*}{\textbf{Model}} & \multirow{2}{*}{\textbf{Setting}} & \multicolumn{4}{c|}{\textbf{Assembly-code Dataset}} & \multicolumn{4}{c}{\textbf{Pseudo-code Dataset}} \\
 & & \textit{Accuracy$\uparrow$} & \textit{F1 Score$\uparrow$} & \textit{FP Rate$\downarrow$} & \textit{Failure Rate$\downarrow$} & \textit{Accuracy$\uparrow$} & \textit{F1 Score$\uparrow$} & \textit{FP Rate$\downarrow$} & \textit{Failure Rate$\downarrow$} \\
\cmidrule{1-10}\morecmidrules\cmidrule{1-10}

\multicolumn{10}{c}{0.5B+ code LLMs} \\
\midrule
\ThreeRowModelFull{Qwen2.5-Coder-0.5B-Instruct}%
{0.497 ,0.355 ,0.260 ,0.104 ,0.437 ,0.312 ,0.241 ,0.196 }%
{0.493 ,0.368 ,0.303 ,0.088 ,0.476 ,0.297 ,0.210 ,0.137 }%
{0.493 ,0.360 ,0.293 ,0.092 ,0.475 ,0.248 ,0.181 ,0.136 }%
\midrule

\multicolumn{10}{c}{1B+ code LLMs} \\
\midrule
\ThreeRowModelFull{DS-Coder-1.3B-Instruct}%
{0.319 ,0.284 ,0.182 ,0.434 ,0.361 ,0.080 ,0.050 ,0.364 }%
{0.402 ,0.300 ,0.217 ,0.276 ,0.429 ,0.314 ,0.244 ,0.210 }%
{0.385 ,0.201 ,0.159 ,0.290 ,0.418 ,0.196 ,0.150 ,0.240 }%

\ThreeRowModelFull{LLM4Decompile-1.3B-v2}%
{0.026 ,0.207 ,0.147 ,0.956 ,0.052 ,0.000 ,0.015 ,0.897 }%
{0.062 ,0.033 ,\cellcolor{gray!90}0.013 ,0.892 ,0.126 ,0.016 ,\cellcolor{gray!90}0.006 ,0.776 }%
{0.079 ,0.048 ,0.020 ,0.857 ,0.119 ,0.017 ,0.007 ,0.786 }%

\ThreeRowModelFull{Qwen2.5-Coder-1.5B-Instruct}%
{0.399 ,0.542 ,0.798 ,0.106 ,0.372 ,\cellcolor{gray!60}0.548 ,0.762 ,0.198 }%
{0.470 ,0.452 ,0.494 ,\cellcolor{gray!40}0.063 ,0.419 ,0.413 ,0.459 ,0.148 }%
{0.459 ,0.471 ,0.544 ,0.067 ,0.442 ,0.482 ,0.491 ,0.147 }%

\ThreeRowModelFull{OpenCoder-1.5B-Instruct}%
{0.465 ,0.218 ,0.179 ,0.144 ,0.468 ,0.144 ,0.078 ,0.179 }%
{0.503 ,0.051 ,\cellcolor{gray!40}0.022 ,0.138 ,0.470 ,0.018 ,0.020 ,0.177 }%
{0.468 ,0.122 ,0.072 ,0.173 ,0.471 ,0.057 ,0.025 ,0.186 }%

\ThreeRowModelFull{Yi-Coder-1.5B-Chat}%
{0.442 ,0.172 ,0.098 ,0.217 ,0.363 ,0.091 ,0.052 ,0.365 }%
{0.495 ,0.048 ,\cellcolor{gray!60}0.015 ,0.153 ,0.454 ,0.014 ,\cellcolor{gray!40}0.009 ,0.209 }%
{0.441 ,0.216 ,0.131 ,0.216 ,0.431 ,0.172 ,0.126 ,0.223 }%
\midrule

\multicolumn{10}{c}{3B+ code LLMs} \\
\midrule
\ThreeRowModelFull{Qwen2.5-Coder-3B-Instruct}%
{0.448 ,0.438 ,0.473 ,0.107 ,0.438 ,0.449 ,0.384 ,0.194 }%
{0.442 ,0.450 ,0.537 ,0.084 ,0.429 ,0.448 ,0.477 ,0.151 }%
{0.466 ,0.464 ,0.492 ,0.083 ,0.468 ,0.520 ,0.470 ,0.151 }%

\ThreeRowModelFull{LLM4Decompile-6.7B-v2}%
{0.242,0.517,0.695,0.478,0.081,0.316,0.413,0.827}%
{0.244,0.526,0.567,0.538,0.157,0.266,0.252,0.705}%
{0.205,0.502,0.727,0.542,0.136,0.128,0.190,0.743}%

\ThreeRowModelFull{Magicoder-S-DS-6.7B}%
{0.356 ,0.474 ,0.522 ,0.297 ,0.284 ,0.314 ,0.263 ,0.464 }%
{0.302 ,0.393 ,0.315 ,0.438 ,0.256 ,0.331 ,0.273 ,0.521 }%
{0.403 ,0.393 ,0.328 ,0.255 ,0.315 ,0.258 ,0.226 ,0.417 }%
\midrule

\multicolumn{10}{c}{7B+ code LLMs} \\
\midrule
\ThreeRowModelFull{Qwen2.5-Coder-7B-Instruct}%
{0.511 ,0.263 ,0.151 ,0.107 ,0.461 ,0.471 ,0.338 ,0.199 }%
{0.515 ,0.280 ,0.171 ,0.093 ,0.458 ,0.446 ,0.358 ,0.170 }%
{\cellcolor{gray!90}0.523 ,0.339 ,0.202 ,0.089 ,0.460 ,0.480 ,0.408 ,0.165 }%

\ThreeRowModelFull{aiXcoder-7B}%
{0.069 ,0.431 ,0.656 ,0.838 ,0.021 ,0.143 ,0.226 ,0.960 }%
{0.138 ,0.456 ,0.527 ,0.715 ,0.090 ,0.194 ,0.089 ,0.844 }%
{0.121 ,0.410 ,0.413 ,0.764 ,0.098 ,0.168 ,0.117 ,0.831 }%

\ThreeRowModelFull{DS-Coder-7B-Instruct-v1.5}%
{0.460 ,0.094 ,0.055 ,0.199 ,0.425 ,0.192 ,0.072 ,0.277 }%
{0.481 ,0.138 ,0.074 ,0.160 ,0.405 ,0.167 ,0.075 ,0.306 }%
{0.488 ,0.140 ,0.050 ,0.167 ,0.401 ,0.133 ,0.067 ,0.308 }%

\ThreeRowModelFull{Starcoder2-7B}%
{0.349 ,0.508 ,0.664 ,0.260 ,0.347 ,0.480 ,0.627 ,0.252 }%
{0.419 ,0.476 ,0.516 ,0.173 ,0.428 ,0.488 ,0.501 ,0.175 }%
{0.434 ,0.427 ,0.433 ,0.158 ,0.429 ,0.475 ,0.501 ,0.156 }%

\ThreeRowModelFull{CodeLlama-7B-Instruct}%
{0.376 ,\cellcolor{gray!60}0.577 ,0.878 ,0.161 ,0.337 ,\cellcolor{gray!40}0.547 ,0.721 ,0.296 }%
{0.381 ,0.514 ,0.758 ,0.132 ,0.401 ,0.509 ,0.642 ,0.160 }%
{0.393 ,0.544 ,0.791 ,0.126 ,0.419 ,0.522 ,0.607 ,0.163 }%

\ThreeRowModelFull{LLMCompiler-7B}%
{0.286 ,0.410 ,0.367 ,0.472 ,0.246 ,0.382 ,0.358 ,0.521 }%
{0.357 ,0.053 ,0.031 ,0.384 ,0.315 ,0.128 ,0.068 ,0.457 }%
{0.373 ,0.057 ,0.026 ,0.361 ,0.318 ,0.093 ,0.047 ,0.462 }%

\ThreeRowModelFull{LLMCompiler-7B-ftd}%
{0.231 ,0.427 ,0.508 ,0.530 ,0.120 ,0.409 ,0.500 ,0.750 }%
{0.264 ,0.177 ,0.075 ,0.564 ,0.206 ,0.178 ,0.103 ,0.624 }%
{0.265 ,0.083 ,0.083 ,0.541 ,0.208 ,0.111 ,0.078 ,0.638 }%

\ThreeRowModelFull{OpenCoder-8B-Instruct}%
{\cellcolor{gray!60}0.522 ,0.172 ,0.106 ,0.076 ,0.481 ,0.239 ,0.180 ,\cellcolor{gray!40}0.116 }%
{0.517 ,0.321 ,0.218 ,0.074 ,0.472 ,0.346 ,0.278 ,0.123 }%
{0.522 ,0.203 ,0.132 ,0.071 ,\cellcolor{gray!40}0.491 ,0.293 ,0.190 ,0.122 }%

\ThreeRowModelFull{Yi-Coder-9B-Chat}%
{0.474 ,0.350 ,0.233 ,0.169 ,0.401 ,0.334 ,0.272 ,0.254 }%
{\cellcolor{gray!40}0.516 ,0.173 ,0.095 ,0.100 ,0.458 ,0.163 ,0.083 ,0.197 }%
{0.477 ,0.409 ,0.377 ,0.098 ,0.432 ,0.355 ,0.300 ,0.190 }%

\ThreeRowModelFull{LLM4Decompile-9B-v2}%
{0.460 ,0.285 ,0.172 ,0.186 ,0.111 ,0.017 ,\cellcolor{gray!60}0.007 ,0.796 }%
{0.462 ,0.251 ,0.188 ,0.160 ,0.270 ,0.093 ,0.069 ,0.496 }%
{0.418 ,0.444 ,0.535 ,0.129 ,0.256 ,0.167 ,0.136 ,0.519 }%
\midrule

\multicolumn{10}{c}{Foundation models} \\
\midrule
\OneRowModel{GPT-3.5-Turbo}{0.492 ,\cellcolor{gray!40}0.568 ,0.725 ,\cellcolor{gray!60}0.001 ,\cellcolor{gray!60}0.502 ,0.546 ,0.639 ,\cellcolor{gray!60}0.006 }%
\OneRowModel{DeepSeek-R1}{0.514 ,\cellcolor{gray!90}0.592 ,0.722 ,\cellcolor{gray!90}0.000 ,\cellcolor{gray!90}0.523 ,\cellcolor{gray!90}0.600 ,0.712 ,\cellcolor{gray!90}0.001 }%
\bottomrule
\end{tabular}
\begin{tablenotes}
\item The top-3 metrics are colored with \colorbox{gray!60}{gray}, where the \colorbox{gray!90}{best}, \colorbox{gray!60}{second-best}, and \colorbox{gray!40}{third-best} metrics display in progressively lighter shades of gray.
\end{tablenotes}
\end{threeparttable}
} 
\end{table*}

\lqy{
\textbf{Overall Results}:
Table~\ref{tab:tab1} presents the performance of 19 open-source code LLMs and two foundation models on the binary SPD dataset, evaluated via three prompting strategies.
All models perform poorly on the dataset with both assembly-code and pseudo-code representations. Specifically, open-source code LLMs exhibit weak instruction-following ability and cannot properly respond with a simple ``yes'' or ``no'' when instructed. We accordingly report an instruction-following failure rate (denoted as “Failure Rate” in Table~\ref{tab:tab1}).
The average failure rate over open-source code LLMs reaches 0.283 and 0.373 on the assembly-code dataset and pseudo-code dataset, respectively. Among them, the LLM4Decompile-1.3B-v2 and the aiXcoder-7B perform the worst, with failure rates as high as 0.956 and 0.960, respectively.
Interestingly, assembly-code representations generally yield slightly better performance than pseudo-code representations. Further analysis reveals that instruction-following tends to degrade on pseudo-code representations. For most failed cases, vanilla code LLMs continue generating pseudo-code instead of responding ``yes'' or ``no''. Consistently, 17 code LLMs exhibit a higher failure rate on pseudo-code representations than on assembly-code representations, except for DS-Coder-1.3B-Instruct and Starcoder2-7B.
}

Based on these results, we derive the following findings.

\lqy{
\textbf{\textit{Off-the-shelf models are not directly suited to binary security patch detection}}. All off-the-shelf code LLMs and even the state-of-the-art foundation models do not reach satisfactory performance on this task.
}

\lqy{
\textbf{\textit{Evaluated prompting strategies provide limited benefits for binary SPD}}. Compared with zero-shot, CoT and self-correction provide improvements in certain models, but remain insufficient for enabling code LLMs to identify security patches effectively. Fundamentally, the lack of domain knowledge of binary SPD within off-the-shelf models cannot be mitigated by prompting strategies.
}

\lqy{
\textbf{\textit{Many off-the-shelf code LLMs exhibit poor instruction-following}}. In particular, LLM4Decompile-1.3B-v2 and aiXcoder-7B exhibit particularly weak instruction-following ability, failing to follow the simple detection instruction. By contrast, instruction-tuned (Instruct) LLMs perform comparatively better. The foundation models and the Qwen2.5 series demonstrate the strongest instruction-following behavior among the evaluated models.
}

\begin{tcolorbox}
[width=\linewidth-2pt,boxrule=3pt,top=3pt, bottom=1pt, left=3pt,right=3pt, colback=gray!10,colframe=gray!10]
\lqy{\textbf{Key findings}:
Off-the-shelf models are not directly suited to binary security patch detection, regardless of the prompting strategy employed. Particularly, many code LLMs that are not instruction-tuned exhibit poor instruction-following ability and are therefore unsuitable for producing detection outputs.
}
\end{tcolorbox}

\subsection{Answer to RQ2: Fine-tuning Yields Substantial Effectiveness for Code LLMs on Binary SPD Tasks}
\textbf{Overall Results}:
Table~\ref{tab:tab2} presents the performance of 19 fine-tuned code LLMs on the binary SPD dataset.
The performance of all code LLMs on the pseudo-code dataset is superior to that on the assembly-code dataset.
On the assembly-code dataset, Qwen2.5-Coder-7B-Instruct achieves the best accuracy and F1 scores of 0.720 and 0.651, and LLMCompiler-7B achieves the lowest false positive rate of 0.189. 
On the pseudo-code dataset, LLM4Decompile-9B-v2 achieves exceptional accuracy, F1 score, and false positive rate of 0.915, 0.897, and 0.058. 

\begin{table*}[htbp]
\centering
\caption{Evaluation results of fine-tuning code LLMs on binary SPD}
\vspace{-1em}
\label{tab:tab2}
\begin{threeparttable}
\begin{tabular}{c|ccc|ccc}
\toprule
\multirow{2}{*}{\textbf{Model}} & \multicolumn{3}{c|}{\textbf{Assembly-code Dataset}} & \multicolumn{3}{c}{\textbf{Pseudo-code Dataset}}\\

& \textit{Accuracy$\uparrow$} & \textit{F1 Score$\uparrow$} & \textit{FP Rate$\downarrow$} & \textit{Accuracy$\uparrow$} & \textit{F1 Score$\uparrow$} & \textit{FP Rate$\downarrow$} \\
\cmidrule{1-7}\morecmidrules\cmidrule{1-7}

\multicolumn{7}{c}{0.5B+ code LLMs} \\
\midrule
Qwen2.5-Coder-0.5B-Instruct & 0.627 & 0.494 & 0.228 & 0.832 & 0.790 & 0.106 \\
\midrule

\multicolumn{7}{c}{1B+ code LLMs} \\
\midrule
DS-Coder-1.3B-Instruct & 0.653 & 0.542 & 0.224 & 0.815 & 0.773 & 0.133 \\
LLM4Decompile-1.3B-v2 & 0.621 & 0.491 & 0.240 & 0.788 & 0.734 & 0.139 \\
Qwen2.5-Coder-1.5B-Instruct & 0.630 & 0.512 & 0.244 & 0.788 & 0.739 & 0.153 \\
OpenCoder-1.5B-Instruct & 0.551 & 0.417 & 0.322 & 0.696 & 0.618 & 0.219 \\
Yi-Coder-1.5B-Chat & 0.627 & 0.524 & 0.269 & 0.752 & 0.691 & 0.179 \\
\midrule

\multicolumn{7}{c}{3B+ code LLMs} \\
\midrule
Qwen2.5-Coder-3B-Instruct & 0.668 & 0.576 & 0.232 & 0.776 & 0.715 & 0.144 \\
LLM4Decompile-6.7B-v2 & 0.610 & 0.502 & 0.283 & 0.804 & 0.756 & 0.131 \\
Magicoder-S-DS-6.7B & 0.597 & 0.472 & 0.278 & 0.756 & 0.704 & 0.194 \\
\midrule

\multicolumn{7}{c}{7B+ code LLMs} \\
\midrule
Qwen2.5-Coder-7B-Instruct & \textbf{0.720} & \textbf{0.651} & 0.207 & 0.877 & 0.851 & 0.085 \\
aiXcoder-7B & 0.621 & 0.495 & 0.246 & 0.867 & 0.836 & 0.083 \\
DS-Coder-7B-Instruct-v1.5 & 0.605 & 0.483 & 0.271 & 0.740 & 0.670 & 0.176 \\
Starcoder2-7B & 0.546 & 0.417 & 0.336 & 0.695 & 0.602 & 0.196 \\
CodeLlama-7B-Instruct & 0.641 & 0.525 & 0.233 & 0.824 & 0.780 & 0.114 \\
LLMCompiler-7B & 0.700 & 0.608 & \textbf{0.189} & 0.901 & 0.879 & 0.062 \\
LLMCompiler-7B-ftd & 0.614 & 0.483 & 0.249 & 0.869 & 0.837 & 0.075 \\
OpenCoder-8B-Instruct & 0.640 & 0.517 & 0.224 & 0.760 & 0.712 & 0.199 \\
Yi-Coder-9B-Chat & 0.637 & 0.531 & 0.251 & 0.828 & 0.786 & 0.112 \\
LLM4Decompile-9B-v2 & 0.681 & 0.595 & 0.225 & \textbf{0.915} & \textbf{0.897} & \textbf{0.058} \\


\bottomrule
\end{tabular}
\begin{tablenotes}
\footnotesize
\item The best metrics are \textbf{bolded}.
\end{tablenotes}
\end{threeparttable}
\vspace{-1em}
\end{table*}

\begin{figure}[h]
\centering
\begin{minipage}[h]{0.75\linewidth}
    \centering
    \includegraphics[width=\textwidth]{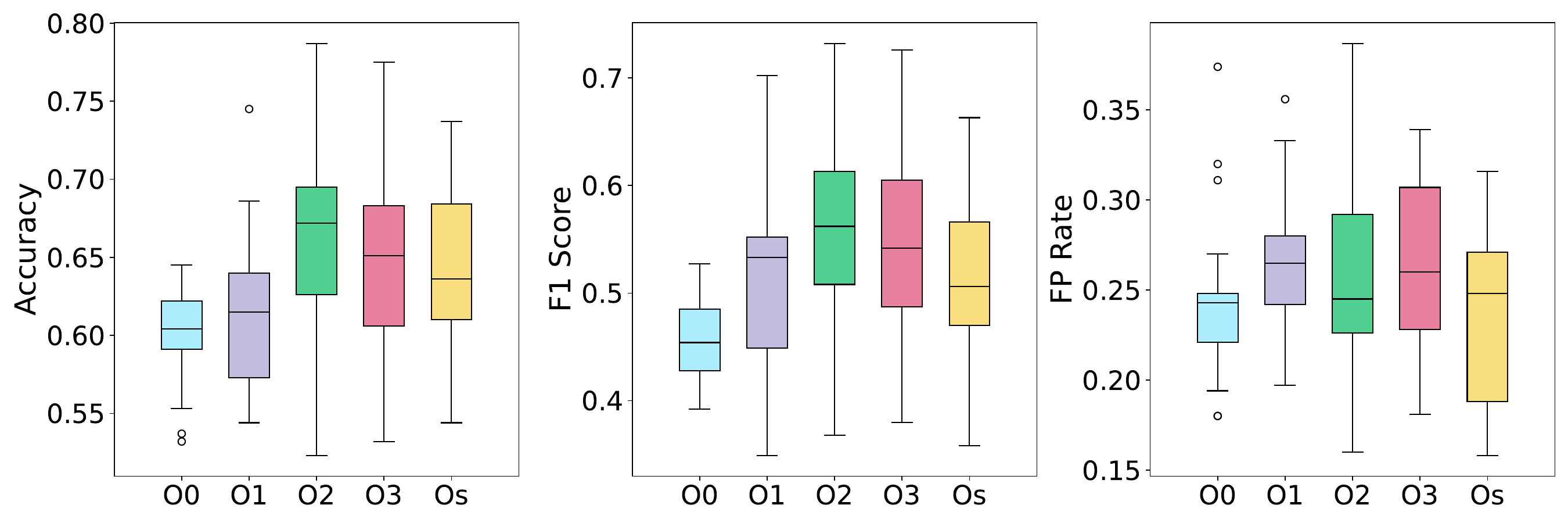}
    \vspace{-2em}
    \caption{Box plots of the metrics on the assembly-code dataset at different optimization levels}
    \label{fig:ass_optim}
\vspace{1em}
\end{minipage}
\centering
\begin{minipage}[b]{0.75\linewidth}
    \centering
    \includegraphics[width=\textwidth]{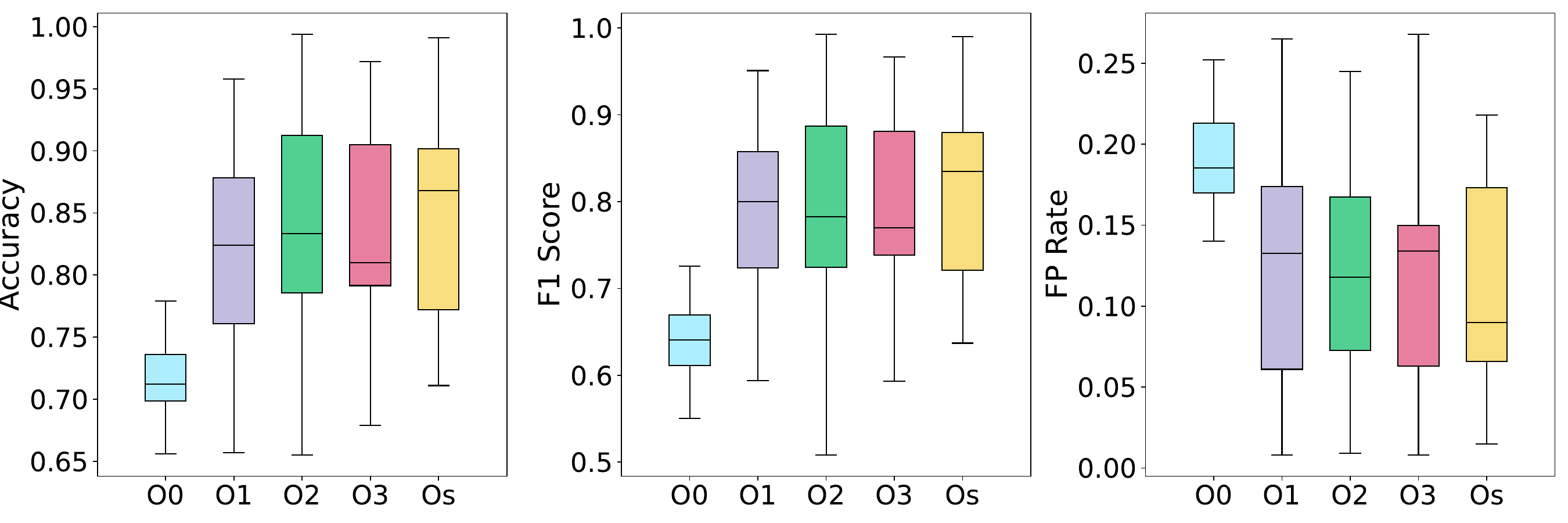}
    \vspace{-2em}
    \caption{Box plots of the metrics on the pseudo-code dataset at different optimization levels}
    \label{fig:pseudo_optim}
\end{minipage}
\vspace{-2em}
\end{figure}

Figures~\ref{fig:ass_optim} and~\ref{fig:pseudo_optim} present the box plots of the three evaluation metrics achieved by 19 code LLMs on both datasets at different optimization levels. 
As shown in Figure~\ref{fig:ass_optim}, code LLMs exhibit insufficient robustness on the assembly-code dataset. In terms of accuracy and F1 score, code LLMs on O0-optimized data exhibit the lowest average value, whereas they achieve the highest average value on O2-optimized data. Moreover, substantial variation exists in Interquartile Ranges (IQRs) across different optimization levels. 
In terms of false positive rate, code LLMs demonstrate a higher average value on O1-optimized data compared to other optimization levels, while more pronounced fluctuations exist in IQRs, especially on O3- and Os-optimized data.
As shown in Figure~\ref{fig:pseudo_optim}, code LLMs exhibit more acceptable robustness on the pseudo-code dataset. 
For O0-optimized data, the differences across the three metrics relative to other optimization level data are substantial, whereas data optimized with O1–Os optimizations exhibits comparable average values and IQRs across all three metrics.

Based on these results, we derive the following findings.

\lqy{
\textbf{\textit{Fine-tuned code LLMs demonstrate effectiveness in binary security patch detection}}.
} 
The binary SPD task is not especially challenging for advanced code LLMs after fine-tuning. Even code LLMs with smaller sizes (e.g., Qwen2.5-Coder-0.5B-Instruct) can effectively learn the domain knowledge of binary SPD through the fine-tuning strategy. They demonstrate well-balanced performance across all three metrics, which are practical in real-world binary SPD tasks. Notably, even the fine-tuned Qwen2.5-Coder-0.5B-Instruct achieves higher accuracy and F1 scores while maintaining a lower false positive rate compared to GPT-3.5-Turbo and DeepSeek-R1.

\lqy{
\textbf{\textit{Fine-tuned code LLMs exhibit scale dependence but do not guarantee statistical monotonicity}}.
}
Code LLMs with 7B parameters generally outperform those with 1B to 3B parameters. Among all the code LLMs, LLM4Decompile-9B-v2, the largest extant model, achieves the highest accuracy and F1 scores.
The Yi-Coder and OpenCoder families exhibit the scaling laws~\cite{kaplan2020scaling}, \zj{whereas DeepSeek-Coder and Qwen2.5-Coder families exhibit non-monotonic scaling behaviors. Notably, DeepSeek-Coder-7B-Instruct-v1.5 underperforms its 1.3B variant, and Qwen2.5-Coder-0.5B-Instruct surprisingly outperforms its larger variants (1.5B, 3B) on pseudo-code while matching 7B-level performance and even surpassing some 8B-9B models (e.g., Yi-Coder-9B-Chat). For resource-constrained deployments, Qwen2.5-Coder-0.5B-Instruct emerges as the optimal efficiency-performance choice. This suggests that task-specific architectural optimizations and training strategies may be more critical than raw parameter count for specialized binary analysis tasks.}

\lqy{
\textbf{\textit{Code LLMs fine-tuned on the pseudo-code dataset achieve superior performance}}.
}
When evaluated solely by accuracy and F1 score, smaller code LLMs fine-tuned on the pseudo-code dataset achieve competitive performance. Specifically, Qwen2.5-Coder-0.5B-Instruct attains accuracy and F1 scores on the pseudo-code dataset that are more than 0.2 higher than those on the assembly-code dataset. However, these smaller models tend to exhibit a higher false positive rate than larger models. Therefore, practical deployment should also consider trade-offs between accuracy and false positive rate.

\lqy{\textbf{\textit{Code LLMs fine-tuned on the pseudo-code dataset exhibit robustness across different optimization levels}}.}
Code LLMs fine-tuned on the pseudo-code dataset exhibit a relatively consistent distribution across different optimization levels in terms of all three metrics. In comparison, Code LLMs fine-tuned on the assembly-code dataset exhibit a marked lack of consistency in the distribution across different optimization levels.

\textit{\textbf{Pre-fine-tuning on low-level code benefits the binary SPD task}}.
The effectiveness of pre-fine-tuning on low-level code (e.g., Assembly and LLVM-IR) is more pronounced in larger-scale code LLMs, while task complexity does not consistently yield benefits.
Specifically, LLMCompiler-7B, pre-fine-tuned on assembly code generation and compiler emulation, outperforms its base model across all metrics, CodeLlama-7B-Instruct.
Similarly, LLM4Decompile-9B-v2, pre-fine-tuned on assembly code and pseudo-code decompilation, achieves measurable improvements in both accuracy and F1 score compared to its base model, Yi-Coder-9B-Chat.
However, LLM4Decompile-1.3B-v2 performs worse than DeepSeek-Coder-1.3B-Instruct in both metrics.
Pre-fine-tuned on the more complex flag tuning task, LLMCompiler-7B-ftd underperforms both LLMCompiler-7B and CodeLlama-7B-Instruct.

\begin{tcolorbox}
[width=\linewidth-2pt,boxrule=3pt,top=3pt, bottom=1pt, left=3pt,right=3pt, colback=gray!10,colframe=gray!10]
\lqy{\textbf{Key findings}:
Code LLMs fine-tuned with the pseudo-code dataset demonstrate exceptional performance in binary SPD tasks. With sufficient resources, larger-scale code LLMs that are pre-fine-tuned on low-level code (e.g., LLM4Decompile-9B-v2 and LLMCompiler-7B) are more advantageous for the binary SPD task.
When resources are constrained, smaller-scale Qwen series models (e.g., Qwen2.5-Coder-0.5B-Instruct) emerge as optimal alternatives, demonstrating particularly outstanding efficiency-performance tradeoffs.}
\end{tcolorbox}

\subsection{Answer to RQ3: Pseudo-code Aligns More Closely with Source Code}
\begin{wrapfigure}{r}{0.5\textwidth}
\centering
\vspace{-1em}
\includegraphics[width=0.9\linewidth]
{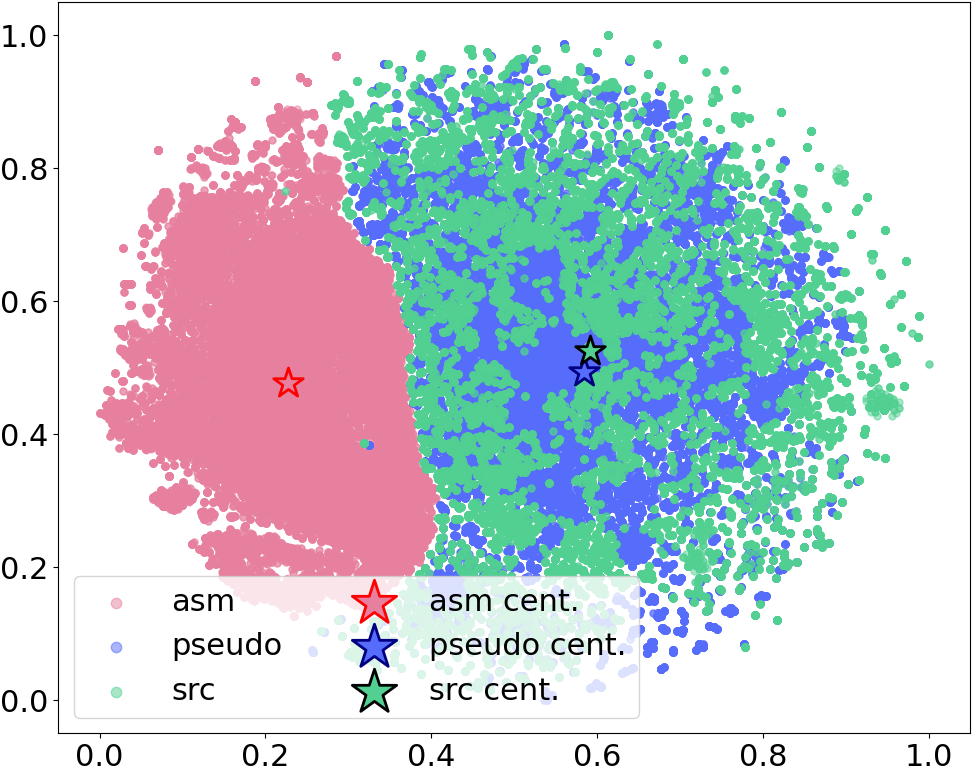}
\vspace{-1em}
\caption{T-SNE plot of different code embeddings}
\label{fig:tsne}

\vspace{2em}

\centering
\includegraphics[width=0.9\linewidth]
{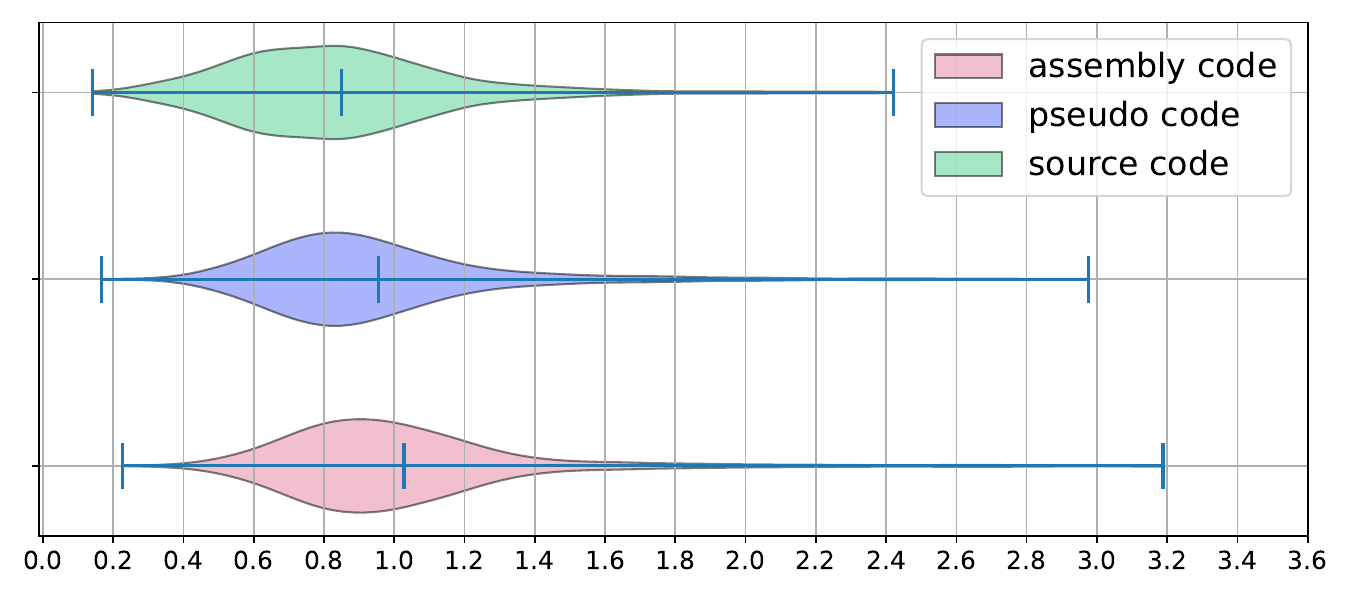}
\vspace{-1em}
\caption{Violin plot of different code naturalness, and samples differing from the mean by more than $4\sigma$ are omitted}
\vspace{-1em}
\label{fig:naturalness}
\end{wrapfigure}

\textbf{Overall Results}:
Figure~\ref{fig:tsne} displays the t-SNE plot for assembly code, pseudo-code, and source code. The centroid distance between source code and pseudo-code is only 0.03, whereas that with assembly code reaches 0.37. Furthermore, the distribution of individual samples in the t-SNE plot indicates that source code and pseudo-code samples essentially overlap; however, a notable separation exists between the source code and the assembly code samples.
Figure~\ref{fig:naturalness} displays the violin plot of code naturalness for assembly code, pseudo-code, and source code. Quantitative analysis reveals that pseudo-code exhibits closer naturalness alignment with source code. The median of source code is 0.823, that of pseudo-code is 0.882, while that of assembly code reaches 0.960. The difference is even more pronounced at the extremum. The maximum naturalness of pseudo-code is 5.323, close to source code’s 4.016, while assembly code exhibits a higher maximum naturalness of 8.028.
\lqy{Both analyses use pre-trained code models as measurement tools, suggesting that the observed similarity patterns are very likely to be reflected in code LLMs.}

Based on these results, we derive the following findings.

\lqy{\textbf{\textit{Pseudo-code is semantically closer to source code}}. The embedding characteristics reveal a stronger similarity between pseudo-code and the corresponding source code than between assembly code and source code, indicating greater semantic proximity in the embedding space.}

\lqy{\textbf{\textit{Pseudo-code is structurally closer to source code and exhibits greater naturalness}}. The code naturalness analysis shows a higher similarity between pseudo-code and the corresponding source code than between assembly code and source code.
Furthermore, pseudo-code consistently achieves lower extreme values, mean, and median in code naturalness than assembly code, indicating that pseudo-code is inherently more natural.}

\lqy{\textbf{\textit{Representations closer to source code are more suited for code LLMs in binary SPD}}. Since pseudo-code consistently aligns more closely with source code across multiple aspects, these results suggest that pseudo-code serves as a more effective input representation for code LLMs on binary SPD tasks.}

\begin{tcolorbox}
[width=\linewidth-2pt,boxrule=3pt,top=3pt, bottom=1pt, left=3pt,right=3pt, colback=gray!10,colframe=gray!10]
\lqy{\textbf{Key Findings}:
Pseudo-code is closer to source code than assembly code. Therefore, it better bridges the gap between the code knowledge embedded in code LLMs and the representations of binary patches. 
In contrast, the significant discrepancy between low-level code (e.g., assembly code) and source code results in insufficient knowledge activation in code LLMs, even when fine-tuned on assembly-code datasets.}
\end{tcolorbox}

\subsection{Answer to RQ4: Augmenting with Source Code Data Enhances Fine-Tuning Effectiveness}
\textbf{Overall Results}:
Table~\ref{tab:tab3} presents the performance of 19 fine-tuned code LLMs on the source-code augmented dataset. 
\lqy{
Experimental results indicate that all 19 code LLMs show improvements in both accuracy and F1.
Notably, LLM4Decompile-1.3B-v2, Qwen2.5-Coder-1.5B-Instruct, OpenCoder-1.5B-Instruct, and Yi-Coder-1.5B-Chat each achieve increases exceeding 0.1 in both accuracy and F1.
Among the evaluated code LLMs, OpenCoder-1.5B-Instruct demonstrates the largest gains, with an accuracy increase of 0.147 and an F1 increase of 0.187.
18 out of 19 code LLMs exhibit a reduction in false positive rate, while the remaining one maintains the same level.
In particular, LLM4Decompile-9B-v2 reduces its false positive rate by 0.021 to 0.037, which is lower than the previously lowest false positive rate of 0.58 observed on the original dataset.
Considering all three metrics, LLM4Decompile-9B-v2 attains the best overall performance on the augmented dataset and also surpasses all models fine-tuned on the original dataset.}
\begin{table}[htbp]
\centering
\vspace{-1em}
\caption{The performance of code LLMs on source-code augmented dataset.}
\label{tab:tab3}
\vspace{-1em}
\resizebox{0.6\linewidth}{!}{
\begin{tabular}{c|ccc}
\toprule
\multirow{2}{*}{\textbf{Model}} & \multicolumn{3}{c}{\textbf{Augmented Dataset}}\\

& \textit{Accuary$\uparrow$} & \textit{F1 Score$\uparrow$} & \textit{FP Rate$\downarrow$} \\
\cmidrule{1-4}\morecmidrules\cmidrule{1-4}

\multicolumn{4}{c}{0.5B+ code LLMs} \\
\midrule
Qwen2.5-Coder-0.5B-Instruct & \graycell{0.893} & \graycell{0.870} & \graycell{0.072} \\ 
\midrule

\multicolumn{4}{c}{1B+ code LLMs} \\
\midrule
DS-Coder-1.3B-Instruct & \graycell{0.869} & \graycell{0.839} & \graycell{0.087} \\
LLM4Decompile-1.3B-v2 & \graycell{0.888} & \graycell{0.865} & \graycell{0.083} \\
Qwen2.5-Coder-1.5B-Instruct & \graycell{0.907} & \graycell{0.889} & \graycell{0.072} \\
OpenCoder-1.5B-Instruct & \graycell{0.843} & \graycell{0.805} & \graycell{0.100} \\
Yi-Coder-1.5B-Chat & \graycell{0.896} & \graycell{0.873} & \graycell{0.065} \\
\midrule

\multicolumn{4}{c}{3B+ code LLMs} \\
\midrule
Qwen2.5-Coder-3B-Instruct & \graycell{0.888} & \graycell{0.865} & \graycell{0.085} \\
LLM4Decompile-6.7B-v2 & \graycell{0.883} & \graycell{0.855} & \graycell{0.068} \\
Magicoder-S-DS-6.7B & \graycell{0.853} & \graycell{0.822} & \graycell{0.108} \\
\midrule

\multicolumn{4}{c}{7B+ code LLMs} \\
\midrule
Qwen2.5-Coder-7B-Instruct & \graycell{0.913} & \graycell{0.895} & \graycell{0.061} \\
aiXcoder-7B & \graycell{0.912} & \graycell{0.893} & \graycell{0.060} \\
DS-Coder-7B-Instruct-v1.5 & \graycell{0.855} & \graycell{0.825} & \graycell{0.110} \\
Starcoder2-7B & \graycell{0.820} & \graycell{0.776} & \graycell{0.121} \\
CodeLlama-7B-Instruct & \graycell{0.902} & \graycell{0.880} & \graycell{0.060} \\
LLMCompiler-7B & \graycell{0.913} & \graycell{0.894} & \graycell{0.051} \\
LLMCompiler-7B-ftd & \graycell{0.916} & \graycell{0.901} & 0.075 \\
OpenCoder-8B-Instruct & \graycell{0.887} & \graycell{0.861} & \graycell{0.069} \\
Yi-Coder-9B-Chat & \graycell{0.921} & \graycell{0.905} & \graycell{0.053} \\
LLM4Decompile-9B-v2 & \graycell{\textbf{0.933}} & \graycell{\textbf{0.919}} & \graycell{\textbf{0.037}} \\
\bottomrule
\end{tabular}
}
\begin{tablenotes}
\footnotesize
\item The metrics showing improvements are highlighted by \graycell{gray} compared to the code LLMs fine-tuned on the pseudo-code dataset.
\end{tablenotes}
\vspace{-1em}
\footnotesize
\end{table}

Based on these results, we derive the following findings.

\lqy{\textbf{\textit{Data augmentation yields larger gains for models with smaller parameter scales}}. 
In particular, the improvements are most pronounced for code LLMs with approximately \textbf{1B} parameters. 
For accuracy, the average increases for code LLMs with different sizes of 0.5B+, 1B+, 3B+, and 7B+ are 0.061, \textbf{0.113}, 0.096, and 0.070. For F1 score, the average increases are 0.080, \textbf{0.143}, 0.119, and 0.090. For false positive rate, the corresponding average changes are +0.029, \textbf{-0.011}, -0.006, and +0.004.
This phenomenon can be attributed to two primary factors. For larger code LLMs, the pseudo-code dataset brings their performance on the binary SPD task close to optimal levels. Additionally, code LLMs adhere to the Scaling Law~\cite{kaplan2020scaling}, which implies that larger models require training on more extensive data to achieve performance improvements comparable to those attained by smaller models.
This finding suggests that, in resource-constrained scenarios, employing small-scale models and augmenting a sufficiently large binary dataset with pseudo-code representations using easily obtainable source code data is a practical approach.}

\begin{tcolorbox}
[width=\linewidth-2pt,boxrule=3pt,top=3pt, bottom=1pt, left=3pt,right=3pt, colback=gray!10,colframe=gray!10]
\lqy{\textbf{Key findings}:
Fine-tuning code LLMs on the augmented dataset further enhances their capabilities, with particularly greater performance gains observed in code LLMs of smaller parameter scales. Notably, employing small-scale models and augmenting a sufficiently large binary dataset with pseudo-code representations using easily obtainable source code data is a practical approach in resource-constrained scenarios.}
\end{tcolorbox}

\section{Error Analysis} \label{sec:error-analysis}
Although code LLMs fine-tuned on the source-code augmented SPD dataset achieve excellent performance, these models still fail to detect some security patches. 
We conduct a statistical error analysis by selecting the Top 5 code LLMs based on accuracy and F1 score, including LLM4Decompile-9B-v2, Yi-Coder-9B-Chat, LLMCompiler-7B-ftd, Qwen2.5-Coder-7B-Instruct, and LLMCompiler-7B to analyze whether their misidentified instances exhibit any correlations.
Among 1248 test data, the five code LLMs miss 56, 60, 51, 65, and 71 security patches, respectively. 

\begin{wrapfigure}{r}{0.4\linewidth}
\centering
\includegraphics[width=0.85\linewidth]{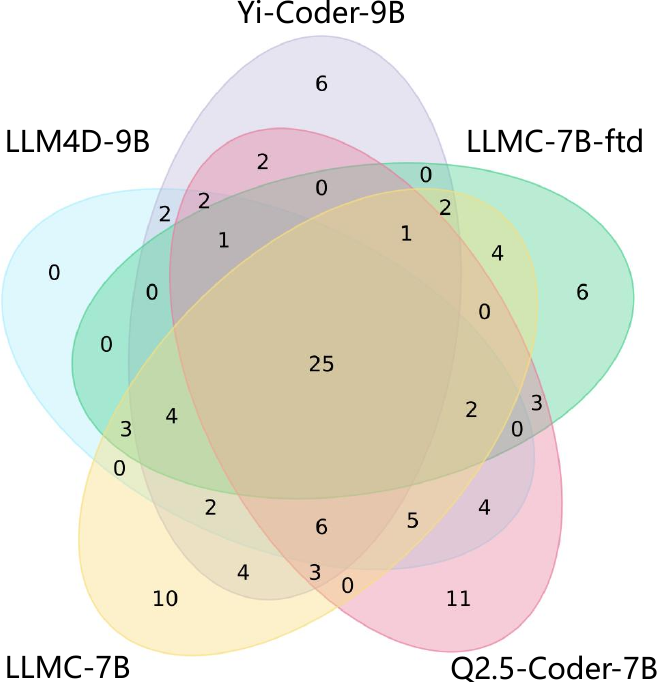}
\caption{Venn diagram of undetected security patches}
\label{fig:error_venn}
\end{wrapfigure}

Figure~\ref{fig:error_venn} illustrates the Venn diagram of the security patches misidentified by these code LLMs.
We analyze these 25 cases individually. Among them, seven patches address denial-of-service (DoS) vulnerabilities (CWE-400); six patches address null pointer dereference vulnerabilities (CWE-476); four patches address buffer overflow vulnerabilities (CWE-119); three patches address out-of-bounds vulnerabilities (CWE-787 and CWE-125); two patches address information exposure vulnerabilities (CWE-200); two patches address improper initialization vulnerabilities (CWE-665); and one patch addresses an integer overflow vulnerability (CWE-190).

We further analyze the patches that are solely misidentified by each of the five code LLMs. Of the five code LLMs, LLM4Decompile-9B-v2 produces none.
Yi-Coder-9B-Chat misidentified six patches: three fix buffer overflow vulnerabilities, one fixes an improper initialization vulnerability, one fixes an out-of-bounds write vulnerability (CWE-787), and one fixes a memory leak vulnerability (CWE-401).
LLMCompiler-7B-ftd misidentified six patches: two fix buffer overflow vulnerabilities, one fixes a null pointer dereference vulnerability, one fixes a race condition vulnerability (CWE-362), one fixes a memory leak vulnerability, and one fixes an information exposure vulnerability.
Qwen2.5-Coder-7B-Instruct misidentified eleven patches: two fix null pointer dereference vulnerabilities, two fix race condition vulnerabilities, two fix improper reference counting vulnerabilities (CWE-911), two fix incorrect calculation vulnerabilities (CWE-682), one fixes an improper locking vulnerability (CWE-667), one fixes a memory leak vulnerability, and one fixes an out-of-bounds write vulnerability. 
LLMCompiler-7B misidentified ten patches: two fix information exposure vulnerabilities, two fix use-after-free (UAF) vulnerabilities (CWE-416), one fixes a double free vulnerability (CWE-415), one fixes a resource management error (CWE-399), one fixes a null pointer dereference vulnerability, one fixes a race condition vulnerability, one fixes an improper reference counting vulnerability, and one fixes an out-of-bounds write vulnerability.

The statistical results indicate that code LLMs generally demonstrate poor abilities in identifying security patches that address memory management issues and the ensuing denial-of-service (DoS) vulnerabilities.
In fact, vulnerabilities related to memory management and dynamic memory issues arising from improper pointer operations represent a major challenge in the vulnerability domain.
In the 2024 Top 25 CWE list\footnote{\url{https://cwe.mitre.org/top25/archive/2024/2024_cwe_top25}}, \textbf{six} distinct vulnerability categories are closely associated with these issues.

\textbf{\textit{Therefore, enabling code LLMs to acquire more domain knowledge regarding vulnerabilities in low-level software (e.g., operating systems) concerning memory management and improper pointer usage is key to addressing the remain challenges in binary SPD.}}

\section{Threats to Validity}
\label{sec:threats-to-validity}

The \textbf{internal threats} are as follows:

\textbf{\textit{Fine-tuning hyperparameter settings}}.
The hyperparameters are determined based on prior research and adjusted manually to mitigate potential threats. Accordingly, we follow the hyperparameter recommendations provided by LLaMAFactory during fine-tuning and set \textit{num\_train\_epochs} to 3 based on validation set performance.

\textbf{\textit{Code large language models selection}}.
Due to GPU limitations, we fine-tune only models with fewer than 13 billion parameters. Consequently, we select as many code LLMs with approximately 7 billion parameters as possible, including nearly all code LLMs below this size.

\textbf{\textit{Difference between closed-source and open-source code}}.
Since security patches for closed-source software are difficult to obtain, and even when available, it is unclear which parts have been fixed. We resort to compiling open-source software as a substitute. To mitigate the discrepancies between closed-source and open-source code, we choose a wide range of popular open-source projects and compile them using different optimization levels to ensure dataset diversity and narrow the gap between the two.

The \textbf{external threats} are as follows:

\textbf{\textit{Non-deterministic output of large language models}}.
The settings for \textit{top\_k} and \textit{temperature} induce non-deterministic outputs in LLMs. Nevertheless, these outputs are not entirely random, as outputs conform to the biases that LLMs learn from extensive training data. We employ the default inference method provided by LLaMAFactory to ensure that the LLM's output remains as stable as possible.
\section{Related Work}
\label{sec:related-work}

\textit{\textbf{Security Patch Detection}}. 
Since patching is inevitable in software systems, the analysis and identification of security patches are crucial for ensuring the security of the software supply chain. In recent years, numerous studies of security patch detection (SPD) have emerged, with learning-based methods achieving the best results. 
PatchRNN~\cite{wang2021patchrnn} applies the Recurrent Neural Network (RNN)~\cite{zaremba2014recurrent} to process both syntactic and semantic features of patches and improve the accuracy of identification security patches in open-source software (OSS). 
GraphSPD~\cite{wang2023graphspd} proposes a graph-based approach, which uses a graph neural network (GNN)~\cite{ggnn} and merges pre-patch and post-patch as code property graphs (CPGs) to effectively capture richer semantics and contextual information from patches, to enhance the precision of security patch detection. 
BinGo~\cite{he2024bingo} explores the use of graph representations to identify security patches in binary code.
They train a GNN to represent subgraphs of binary patches to identify security patches. 
LLMDA~\cite{tang2023just} leverages the capabilities of large language models by jointly representing patches and related texts (e.g., patch explanation and description) to improve performance.
Most recently, RepoSPD~\cite{wen2024repository} considers the repository-level dependencies of patches by constructing a novel graph structure, RepoCPG, and employs a structure-aware patch representation that fuses both the graph and sequence branches, achieving the state-of-the-art. 
Despite the plenty of studies on SPD in source code, none have explored the potential of utilizing the comprehensive code-specific knowledge embedded in code LLMs and the similarity between pseudo and source code to investigate the capability of code LLMs in binary code SPD tasks.

\textit{\textbf{Vulnerability Repair}}.
The target of security patch detection is identifying security patches, while the target of vulnerability repair is generating security patches. Over the past years, learning-based Automated Vulnerability Repair (AVR) has become a promising approach to address vulnerability patching.
VulRepair~\cite{fu2022vulrepair} is a vulnerability repair method that leverages the pre-trained model. By fine-tuning CodeT5~\cite{wang2021codet5} on vulnerability-fix pairs, it achieves excellent repair results.
VRepair~\cite{chen2022neural} utilizes transfer learning by first pre-training on a bug-fix dataset and then further fine-tuning on vulnerability-fix pairs to improve its ability to repair vulnerable C functions.
VulMaster~\cite{zhou2024out} alleviates the sequence-length limitation of Transformer using FiD and applies multitask learning to joint vulnerability prediction and repair tasks, enhancing the repair capability of CodeT5.
VulAdvisor~\cite{zhang2024vuladvisor} fine-tunes CodeT5 using ChatGPT-enhanced data for vulnerability suggestion and demonstrates that VulAdvisor improves the repair capabilities of both AVR methods and developers.

\textit{\textbf{Fine-tuning LLMs for Code-related Tasks}}.
Recently, fine-tuning LLMs has become one of the most common methods for handling code-related tasks.
RepairLlama~\cite{silva2023repairllama} employs Parameter-Efficient Fine-Tuning (PEFT)~\cite{ding2023parameter} to adapt CodeLlama-7B~\cite{roziere2023code} for program repair tasks, achieving superior performance across three benchmarks.
\citet{weyssow2023exploring} verify the superiority of PEFT over ICL and RAG across a diverse set of LLMs for code generation tasks while maintaining reasonable resource consumption.
SteloCoder~\cite{pan2023stelocoder} fine-tunes StarCoder~\cite{listarcoder} applying Low-Rank Adaptation (LoRA)~\cite{hu2022lora} for the multi-programming language-to-Python code translation task ahead of the leaderboard.
\section{Conclusion}
\label{sec:clucusion}
\lqy{
In this paper, we systematically investigate the capabilities of code large language models (LLMs) for binary security patch detection (SPD).
Given the absence of existing datasets for binary SPD, we first construct an open-source, large-scale, multi-project binary SPD dataset that includes multiple optimization levels.
We subsequently evaluate three prompting strategies, including zero-shot, Chain-of-Thought (CoT), and Self-correction, across 19 off-the-shelf code LLMs and two foundation models. Experimental results demonstrate that vanilla models perform poorly on this task. Although CoT and Self-correction improve the performance for certain models compared with zero-shot, they still remain insufficient for enabling code LLMs to identify security patches. It suggests that these prompting strategies cannot mitigate the lack of domain knowledge of binary SPD within vanilla models.
In light of directly prompting being insufficient, we investigate whether fine-tuning is effective. We fine-tune the 19 open-source code LLMs using Low-Rank Adaptation (LoRA). Experimental results demonstrate that fine-tuning substantially improves model performance on binary SPD. Furthermore, models fine-tuned on the pseudo-code dataset consistently outperform those fine-tuned on the assembly-code dataset.
To investigate why fine-tuned code LLMs achieve superior performance on pseudo-code representations, we quantify the similarity between pseudo-code, assembly code, and source code. Experimental results demonstrate that pseudo-code is closer to source code than assembly code, both in embedding characteristics and code naturalness.
Motivated by this finding, we investigate whether augmenting the pseudo-code dataset with source code further improves fine-tuning. Experimental results demonstrate that fine-tuning on the augmented dataset yields measurable gains, with the improvements being particularly pronounced for small-scale models. Thus, under resource-limited conditions, augmenting a sufficiently large pseudo-code dataset with easily obtainable source code emerges as a practical and effective strategy of enhancing the fine-tuning performance of small-scale models in binary SPD.
}

\lqy{
In summary, our empirical study provides an instructive understanding of how code LLMs can be leveraged for binary SPD tasks, highlighting pseudo-code as the most suited data representation.
}


\end{document}